\documentclass[prb,twocolumn,floatfix,superscriptaddress,showpacs]{revtex4}
\usepackage{graphicx}
\usepackage{amsmath,amssymb,bm}
\DeclareMathOperator{\sgn}{sgn}

\begin{document}
\title{A novel approach to transport
  through correlated quantum dots}
\author{C.\ Karrasch}
\affiliation{Institut f\"ur Theoretische Physik, Universit\"at G\"ottingen, 
Friedrich-Hund-Platz 1, D-37077 G\"ottingen, Germany}
\author{T.\ Enss}
\affiliation{Istituto Nazionale di Fisica della Materia--SMC--CNR and
  Dipartimento di Fisica, Universit\`a di Roma ``La Sapienza'', 
  Piazzale Aldo Moro 2, I-00185 Roma, Italy}
\author{V.\ Meden}
\affiliation{Institut f\"ur Theoretische Physik, Universit\"at G\"ottingen, 
Friedrich-Hund-Platz 1, D-37077 G\"ottingen, Germany}

\begin{abstract}
We investigate the effect of local Coulomb correlations on 
electronic transport through a variety of coupled quantum dot 
systems connected to Fermi liquid leads.  We use 
a newly developed functional renormalization group 
scheme to compute the  gate voltage dependence of the linear
conductance, the transmission phase, and the dot occupancies. A detailed
derivation of the flow equations for the dot level positions, 
the inter-dot hybridizations, and the effective 
interaction is presented. For specific setups and parameter sets 
we compare the results to existing accurate numerical 
renormalization group data. This shows that our approach covers the 
essential physics and is quantitatively correct up to fairly large 
Coulomb interactions while being much faster, very flexible, and 
simple to implement. We then demonstrate the power of our method 
to uncover interesting new physics. In several dots coupled in series 
the combined effect of correlations and asymmetry leads to a vanishing
of transmission resonances. In contrast, for a parallel double-dot 
we find parameter regimes in which the two-particle interaction 
generates additional resonances.
\end{abstract}
\maketitle     

\section{Introduction}
\label{intro}

Electronic transport through multi-level quantum dots and several 
coupled dots is currently of great experimental and theoretical
interest due to the possible application of such systems in 
interferometers and for charge- and spin-based quantum
information processing.\cite{Sohn,Loss,Florian,Holleitner1,Chen,Sigrist,Holleitner2,Petta1,Craig,Koppens,Petta2,Johnson} 
The smallness of the mesoscopically confined electron droplet---the 
quantum dot---leads to fairly large energy level spacings and at 
sufficiently small temperatures $T$ only a few levels must be
considered. The physics is strongly affected by charging 
effects\cite{Sohn} (Coulomb blockade) and local Coulomb correlations, 
e.g.\ leading to the Kondo 
effect.\cite{Hewson,Glazman,Ng,Goldhaber,Wiel,Glazmanrev} 

We here present a new method to describe the low-temperature transport 
through both a single multi-level dot and several coupled dots. To
keep the notation short when  referring to both situations in the
following we denote the dot(s) as the mesoscopic system. 
We investigate the linear response conductance $G$ through the setup
as a function of a  gate voltage $V_g$ that shifts the energy levels. 
The connection between the mesoscopic system and the leads 
(the reservoirs) is modeled by low-transmission tunneling barriers. 

\begin{figure*}[htb]
\begin{center}
\includegraphics[width=0.8\textwidth,clip]{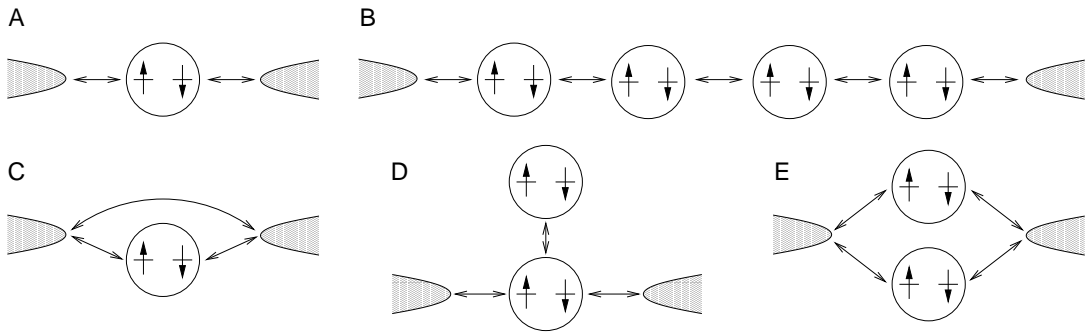}
\end{center}
\vspace{-0.6cm}
\caption[]{Dot geometries considered in the present paper.
\label{fig1}}
\end{figure*}

The transport through a single-level dot with spin degeneracy (see
Fig.~\ref{fig1} A) is well understood. For simplicity we only consider equal 
couplings to the left and right lead. At small $T$ and for 
noninteracting dot electrons $G(V_g)$ shows a Lorentzian resonance 
of unitary height $2 e^2/h$ placed roughly at the energy of the 
single-particle level of the dot. The full width $2 \Gamma$ of the resonance sets 
an energy scale $\Gamma$ that is associated with the strength of the 
tunneling barriers. Including a Coulomb interaction $U$ between the 
spin up and down dot electrons the line shape is substantially
altered as can be seen from the exact $T=0$ Bethe ansatz
solution.\cite{Tsvelik} 
For increasing $U/\Gamma$ it is gradually transformed into a 
box-shaped resonance of unitary heights with a plateau of width $U$ 
and a sharp decrease of $G$ to the left and right of 
it.\cite{Glazman,Ng,Theo1,Gerland} For gate voltages within the
plateau the dot is half-filled implying a local spin-1/2 degree of 
freedom on the dot. Thus, the Kondo effect\cite{Hewson} 
leads to resonant transport throughout this Kondo regime. 
A linear chain of a few coupled single-level dots (see Fig.~\ref{fig1} B)
presents a simple extension of the single-dot case and was studied 
recently.\cite{Oguri2,Oguri3,Zitko2}  This system can equally be
viewed as a short Hubbard chain. 

In more complex cases in which electrons can pass the mesoscopic
system following different paths, not only correlations but also quantum 
interference plays an important role, leading e.g.\ to the Fano\cite{Fano}
effect. Realizations of such a situation that have been studied
theoretically are: 
i) A single dot coupled to two leads that are in addition
coupled by a direct transmission channel (Aharonov-Bohm interferometer with an
embedded dot) as sketched in Fig.~\ref{fig1} C.\cite{Hofstetterkondofano} 
ii) A system of two dots, one directly connected to the two leads and 
the other side-coupled to the first one\cite{Kim1,Cornaglia,Zitko} shown in
Fig.~\ref{fig1} D. iii) Parallel double dots coupled to common 
leads\cite{Izumida,Boese1,Boese2,cir} (see Fig.~\ref{fig1} E) that have also been realized
experimentally.\cite{Holleitner1,Chen,Sigrist,Holleitner2,Petta1,Koppens,Petta2}
The combined effect of multi-path interference through several nearly 
degenerate levels of a single dot and correlations is discussed 
as one of the possible reasons for the puzzling behavior of the 
transmission phase measured in an Aharonov-Bohm 
interferometer.\cite{HeiblumExps} Without a direct hopping between the
dots the setup in Fig.~\ref{fig1} E can either be viewed as two dots each having a
single level or a single dot with two levels. For certain parameter
sets the cases  Fig.~\ref{fig1} D and E are equivalent (see below).\cite{Boese2} 

The numerical renormalization group (NRG) is a reliable method to
investigate physical properties of systems with local Coulomb
correlations\cite{Wilson,Krishnamurthy} and was very successfully 
used to study the linear transport through a single-level 
dot.\cite{Theo1,Gerland,Hofstetterkondofano} Unfortunately, the 
computational resources required strongly increase with the number of 
interacting degrees of freedom. This usually limits the
applicability of NRG to mesoscopic systems with two spin-degenerate
levels\cite{Izumida,Boese1,Boese2,Cornaglia,Zitko,cir} and even in these cases 
a complete analysis of the parameter dependencies is practically
impossible. For a linear chain of dots using special symmetries and 
a direct relation between the linear conductance and phase shifts it was
possible to compute $G$ at $T=0$ for up to four dots.\cite{Oguri2,Oguri3}
 
Thus, alternative reliable methods are needed to
systematically investigate parameter dependencies and study more
complex systems with a larger number of degrees of freedom. 
Simple methods such as low-order perturbation theory or the 
self-consistent Hartree-Fock approximation (SCHFA)
already fail to correctly describe the Kondo physics\cite{Hewson} 
in a single-level dot and are thus inappropriate. 
Equations-of-motion techniques were
successfully used to compute $G(V_g)$ at high temperatures, but
even sophisticated decoupling schemes only qualitatively capture the
small-$T$ Kondo regime.\cite{Kashcheyevs} Similarly, perturbation
theory in the coupling of the mesoscopic system to the leads works
at large $T$, but fails for the large conductance
resonances at small $T$.\cite{Oreg} 
In temperature and parameter regimes in which charge fluctuations dominate 
a real-time renormalization group method\cite{Herbert} can be 
used.\cite{Boese1} 
A zero-temperature method
that has frequently been used is exact diagonalization of a small
cluster of lattice sites containing the dot followed by an embedding
procedure.\cite{Buesser} It has been 
criticized\cite{Ogurikritik,Cornaglia} to produce severe artifacts in 
$G(V_g)$ and even in the case of a single-level dot, where at least 
the qualitative behavior comes out correctly, the quantitative
agreement with NRG data is rather poor.\cite{Torio} 
Also the mean-field
slave-Boson approach\cite{Kotliar} was used, first to study the
single-level dot. Within this method $G(V_g)$
falls off too quickly away from the plateau region as 
compared to the Bethe ansatz and NRG data.\cite{Dong1,Takahashi}   
Furthermore, the method does not capture the correct 
dependence of $G(V_g)$ on a magnetic field that lifts the 
spin degeneracy of the single level.\cite{Dong2} Despite these 
shortcomings the method was also used for more complex dot
systems.\cite{Tanaka} 
Recently, a method based on the
Gunnarsson-Sch\"onhammer variational approach\cite{GS} has been
applied to the dot with a spin-degenerate
level\cite{Rejec} and a linear chain of three dots.\cite{Zitko2}
It remains to be seen if this approach can with
equal success be used in the presence of a magnetic field and for 
multi-path systems. 

We here propose a novel $T=0$ approach to study transport 
through mesoscopic
systems with local Coulomb correlations that is based on the
functional renormalization group (fRG).  The fRG was recently 
introduced\cite{Manfred} as a new powerful tool for studying
interacting Fermi systems. It provides a systematic
way of resumming competing instabilities\cite{2dsystems} and goes
beyond simple perturbation theory even in problems which are not 
plagued by infrared divergences.\cite{Ralf,Hon} 
The fRG procedure 
we use starts from an exact hierarchy of differential flow
equations for the one-particle irreducible vertex
functions,\cite{Wetterich,Morris,SalmhoferHonerkamp,Ralf} 
i.e., the self-energy, the irreducible 2-particle interaction etc.  
It is derived by replacing the free propagator by a propagator depending on
an infrared cutoff $\Lambda$ and taking the derivative of the generating 
functional with respect to $\Lambda$. We will apply two different
truncation schemes to the models sketched in Fig.~\ref{fig1}. Both
result in a simple set of coupled differential equations that can easily be
integrated numerically and in certain cases even analytically. By an
explicit comparison of our results for $G(V_g)$ and the occupancies 
to NRG and Bethe ansatz data we show
that the method works  well for surprisingly large $U/\Gamma$.
The method can easily be extended to other setups and thus 
provides a useful and reliable tool to
investigate transport through quantum dots. In two short publications
it was earlier used to compute $G(V_g)$ for a single dot
with Luttinger-liquid leads\cite{STV} and for a spin-polarized  parallel
double dot.\cite{cir} 

This paper is organized as follows. In Sec.~\ref{model} we present our
general model. We then derive the fRG flow equations  
in Sec.~\ref{method}. In Secs.~\ref{singledot}-\ref{paralleldoubledot}
we discuss our results for the gate voltage dependence of the linear
conductance of the systems of Fig.~\ref{fig1} and for the cases shown
in A-D we explicitly compare them to existing NRG and Bethe ansatz 
data. For each situation we judge the quality of our
approximate method and discuss the limits of its applicability. 
In Secs.~\ref{lindot}-\ref{paralleldoubledot} (setup of Fig.~\ref{fig1} B, D, and 
E) we go beyond situations that have previously been studied and 
present novel results. 
We conclude in Sec.~\ref{summary} with a brief summary and an outlook.

\section{The model}
\label{model}

Our general Hamiltonian 
\begin{eqnarray}
\label{ham}
H = H_{\text{lead}} + H_{\text{direct}} + H_{\text{dot}} + H_{\text{coup}}
\end{eqnarray}
consists of four parts. 
For simplicity the two leads are assumed to be equal and described by a 
tight-binding model 
\begin{eqnarray}
\label{leadham}
H_{\text{lead}} = -\tau  \sum_{\sigma} \sum_{l=L,R} \sum_{m=0}^{\infty} 
\left( c^\dag_{m,\sigma,l} c_{m+1,\sigma,l} + \text{H.c.} \right) 
\end{eqnarray}
with $c^{(\dag)}_{m,\sigma,l}$ being the annihilation (creation)
operator for an electron with spin direction
$\sigma=\uparrow,\downarrow$ localized on lattice site $m$ of the left
($l=L$) or right ($l=R$) lead. The hopping matrix element in the leads
is denoted by $\tau$. 
To describe the system sketched in Fig.~\ref{fig1} C we need a term that directly 
links the left and right lead
\begin{eqnarray}
\label{directham}
 H_{\text{direct}} = \sum_{\sigma} 
   \left(  t_{\text{LR}} c^\dag_{0,\sigma,R} c_{0,\sigma,L} + \text{H.c.}  \right) \; .
\end{eqnarray} 
The coupled quantum dots are modeled as
\begin{eqnarray}
\label{dotham}
&& H_{\text{dot}} =  \sum_{\sigma} \sum_{j} \varepsilon_{j,\sigma} 
d_{j,\sigma}^\dag d_{j,\sigma} \nonumber \\ &&
-  \sum_{\sigma} \sum_{j > j'} t_{j,j'} d_{j,\sigma}^\dag d_{j',\sigma} 
+  \text{H.c.} \nonumber \\ && 
+  \frac{1}{2} \; \sum_{\sigma,\sigma'} \sum_{j,j'} U_{j,j'}^{\sigma,\sigma'} 
\left(n_{j,\sigma}-\frac{1}{2}\right)
\left(n_{j',\sigma'}-\frac{1}{2}\right)  
\end{eqnarray} 
with annihilation (creation) operators $d_{j,\sigma}^{(\dag)}$
and the site occupancy $n_{j,\sigma}=d_{j,\sigma}^{\dag}d_{j,\sigma}$.
With this dot Hamiltonian we equally describe multi-level and 
multi-dot situations or even mixtures of both. The dot or level index 
$j$ runs from $1$ to $N$. 
The level positions $\varepsilon_{j,\sigma}$  
consist of constant parts $\varepsilon_{j,\sigma}^0$ and 
a variable gate voltage $V_g$ that is used to
shift the energy levels: $\varepsilon_{j,\sigma} =
\varepsilon_{j,\sigma}^0 + V_g$. 
Our notation allows for both spin-degenerate
levels and levels that are split by a magnetic field. 
For $j$'s that belong to different dots it 
is meaningful to introduce nonzero hopping matrix elements $t_{j,j'}$. 
The 2-particle interaction $U_{j,j'}^{\sigma,\sigma'}$ vanishes 
for $j=j'$ and $\sigma =\sigma'$. 
Besides this restriction we allow for both intra-level 
interactions of spin up and spin down electrons as well 
as inter-level interactions. 
Finally, the coupling between dot and lead states is given by 
\begin{eqnarray}
\label{dotleadham}
H_{\text{coup}} = - \sum_{\sigma} \sum_{l=L,R} \sum_{j} (t_j^l 
c^\dag_{0,\sigma,l} d_{j,\sigma} + \text{H.c.})
\end{eqnarray}  
with overlaps $t_j^l$, that enter the energy scale of the dot level 
broadening $\Gamma_j^l = \pi |t_j^l|^2 \rho_{\text{lead}}$, 
where $\rho_{\text{lead}}$ denotes the local density of states at the end of 
each semi-infinite lead. As usual we later take $\rho_{\text{lead}}$ to be 
energy independent (wide-band limit). 
The choice of the nonzero $t_j^l$ and $t_{j,j'}$ introduces a
spatial geometry between the dots. For ring-like geometries 
(Fig.~\ref{fig1} C and E) one might also be interested in studying the 
effect of a magnetic flux piercing the ring. This can be modeled 
by including appropriate phase factors in the hopping matrix elements 
$t_{\text{LR}}$, $t_{j}^l$, $t_{j,j'}$ along the 
ring.\cite{Hofstetterkondofano,Boese1,cir}   
All the systems shown in Fig.~\ref{fig1} follow from this 
Hamiltonian by special choices of the parameters. Obviously, it can 
also be used to model other cases of interest. 

\section{The method}
\label{method}

Applying the Landauer-B\"uttiker formalism\cite{Datta} the $T=0$ linear
conductance for noninteracting electrons can be expressed as
\begin{eqnarray}
\label{LB1}
G(V_g) = G_\uparrow(V_g) + G_\downarrow(V_g)  \; ,
\end{eqnarray} 
with
\begin{eqnarray}
\label{LB2}
G_\sigma(V_g) = \frac{e^2}{h} \, \left| {\mathcal T}_\sigma(0,V_g) \right|^2 \; ,
\end{eqnarray} 
where the transmission is given by 
\begin{eqnarray}
\label{LB3}
 {\mathcal T}_\sigma(\varepsilon,V_g)  =  2 \pi \tau^2 \; \rho_{\text{lead}}
 (\varepsilon) \; 
   {\mathcal G}_{0,\sigma,L;0,\sigma,R}(\varepsilon+ i0)  
\end{eqnarray} 
and the energy $\varepsilon$ is measured relative to the chemical 
potential $\mu$. Here $\mathcal G$ denotes the retarded 
one-particle Green function
(the resolvent in the present single-particle problem) of the 
system. For the leads described by Eq.~(\ref{leadham}) the local
density of states at site $m=0$ is (if they are decoupled from the
rest of the system) 
\begin{eqnarray}
\label{randDOS}
\rho_{\text{lead}}(\varepsilon) = \frac{1}{2\pi \, \tau^2} \; \sqrt{4
  \tau^2-(\varepsilon + \mu)^2}  \;  \;  
\Theta( 2 \tau - |\varepsilon + \mu|)\; .
\end{eqnarray}    
The transmission phase $\alpha_\sigma(V_g)$ is given by the phase of $ {\mathcal
  T}_\sigma(0,V_g)$. 

The noninteracting Green function matrix element involving lead states 
${\mathcal G}_{0,\sigma,L;0,\sigma,R}$ can be related to 
matrix elements involving only the dot states with quantum numbers
$j,\sigma$ and $\rho_{\text{lead}}$. 
The explicit form of this relation depends on the 
choice of nonzero $t_j^l$. It
can be obtained using a standard projection technique\cite{Taylor}   
and is based on the relation 
\begin{eqnarray}
\label{projection}
P{\mathcal G}_{1p}(z)P  & = & \\ 
&& \hspace{-2.0cm}\left [ zP-PH_{1p}P-PH_{1p}Q\left 
(zQ-QH_{1p}Q \right )^{-1}QH_{1p}P\right ]^{-1} \, , \nonumber
\end{eqnarray}
where $H_{1p}$ is an arbitrary single-particle Hamiltonian, ${\mathcal
  G}_{1p}(z)=\left[ z -  H_{1p} \right]^{-1}$ the resolvent, and $P$,
$Q$ are projectors with $P+Q=\mathbf 1$. 

For Fermi liquids, as discussed here, the same $T=0$ relations 
Eqs.~(\ref{LB1})-(\ref{LB3}) between matrix elements of the 
one-particle Green function and the conductance hold in the presence 
of interactions.\cite{Oguri1}  
To obtain $G_\sigma$ and $\alpha_\sigma$ we thus have to determine 
the one-particle Green function of the dot system in the 
presence of the interaction 
and the leads. From this also the level occupancies $\left< n_{j,\sigma} 
\right>$ can be determined by integrating 
${\mathcal G}_{j,\sigma;j,\sigma}(z)$ along the imaginary 
axis.\cite{NegeleOrland}  
 
As a first step we integrate out the noninteracting leads within a
functional integral representation of our  many-body
problem.\cite{NegeleOrland} They provide a frequency dependent  
one-particle potential for those indices $j$ for which $t_j^l \neq 0$. 
On the imaginary frequency axis it is given by 
\begin{eqnarray}
\label{leadpotdef}
V^{\text{lead}}_{j,\sigma;j',\sigma'}(i\omega) = \sum_{l} t_j^l  \left(t_{j'}^l\right)^\ast  
\text{g}_{\text{lead}} (i\omega)  \, \delta_{\sigma,\sigma'} \;,
\end{eqnarray}
where $\text{g}_{\text{lead}}(i \omega)$ denotes the spin-independent 
Green function of the isolated semi-infinite leads taken at the 
last lattice site
\begin{eqnarray}
\label{gleaddef}
\text{g}_{\text{lead}}(i \omega) = \frac{i\omega+\mu}{2 \; \tau^2} \left( 1 - 
 \sqrt{1 - \frac{4 \; \tau^2}{(i\omega+\mu)^2}} \, \right) \; .
\end{eqnarray}
After this
step, instead of dealing with an infinite system we only have to 
consider the dot system of $N$ interacting levels. We here take 
$\mu=0$. In the computation of the
Green function projected onto the dot system the sum of the 
dot Hamiltonian at $U_{j,j'}^{\sigma,\sigma'} = 0$ 
and $V_{j,j'}^{\text{lead}}(i\omega)$ can be interpreted as a frequency 
dependent ``effective Hamiltonian'' and in the following 
will be denoted by $h(i\omega)$.

We now set up our fRG scheme.  
As a first step we replace the noninteracting dot
propagator ${\mathcal G}_0$, as
obtained from Eq.~(\ref{dotham}) and the projection of the leads, by a cutoff-dependent 
propagator that suppresses the infrared modes with Matsubara frequency $|\omega| < \Lambda$,
\begin{equation}
\label{cutoffproc}
{\mathcal G}_0^{\Lambda}(i \omega) = \Theta(|\omega|-\Lambda) 
{\mathcal G}_0(i \omega) =  \Theta(|\omega|-\Lambda)  \left[ i\omega -
h(i \omega) \right]^{-1}
\end{equation}
with $\Lambda$ running from $\infty$ down to $0$. For convenience we
use a sharp cutoff (see below). Using 
${\mathcal G}_0^\Lambda$ in the 
generating functional of the irreducible vertex 
functions\cite{NegeleOrland} and taking the derivative with respect to 
$\Lambda$ one can derive an exact, infinite hierarchy of coupled differential
equations for vertex functions, such as the self-energy and the
irreducible 2-particle interaction. In 
particular, the flow of the self-energy $\Sigma^\Lambda$ (1-particle
vertex) is determined by $\Sigma^\Lambda$ and 
the 2-particle vertex $\Gamma^{\Lambda}$, while the flow of 
$\Gamma^{\Lambda}$ is determined by $\Sigma^\Lambda$, $\Gamma^{\Lambda}$, and 
the flowing 3-particle vertex $\Gamma_3^{\Lambda}$.
The latter could be computed from a flow equation involving
the 4-particle vertex, and so on.
At the end of the fRG flow $\Sigma^{\Lambda=0}$ is the self-energy $\Sigma$ 
of the original, cutoff-free problem we are interested in.\cite{SalmhoferHonerkamp,EnssThesis,Ralf}
A detailed derivation of the fRG flow equations for a general quantum
many-body problem that only requires a basic knowledge of the
functional integral approach to many-particle
physics\cite{NegeleOrland} and the application of the method for a
simple toy problem are presented in Ref.~\onlinecite{lecturenotes}.   
In practical applications the hierarchy of flow equations has to be truncated and
$\Sigma^{\Lambda=0}$ only provides an approximation for the exact 
$\Sigma$. As a first approximation we here neglect the 3-particle vertex
(irreducible 3-particle interaction).
The contribution of $\Gamma_3^{\Lambda}$ to $\Gamma^{\Lambda}$ is small
as long as $\Gamma^{\Lambda}$ is small, because 
$\Gamma_3^{\Lambda}$ is initially (at $\Lambda=\infty$) zero 
and is generated only from terms of third order in 
$\Gamma^{\Lambda}$. Furthermore,  $\Gamma^{\Lambda}$ stays small 
for all $\Lambda$ if its initial values $U_{j,j'}^{\sigma,\sigma'}$ 
(for the details of the initial conditions, see below) are not too large. 
By explicit comparison to NRG data below we will clarify 
the meaning of ``not-too-large'' 
in the cases of interest. This approximation leads to a closed set of 
equations for $\Gamma^{\Lambda}$  and $\Sigma^{\Lambda}$ 
given by
\begin{equation}
\label{flowS}
 \frac{\partial}{\partial\Lambda} \Sigma^{\Lambda}(1',1) =
 \, - \, \frac{1}{2 \pi} \, \sum_{2,2'} \, e^{i \omega_{2} 0^+} \,
 S^{\Lambda}(2,2') \; \Gamma^{\Lambda}(1',2';1,2) 
\end{equation}
and
\begin{eqnarray}
 && \mbox{} \hspace{-1.cm} \frac{\partial}{\partial\Lambda} \Gamma^{\Lambda}(1',2';1,2)
 = \; \frac{1}{2 \pi} \,
 \sum_{3,3'} \sum_{4,4'} \, {\mathcal G}^{\Lambda}(3,3') \, S^{\Lambda}(4,4')
 \nonumber \\
 &&  \mbox{} \hspace{-1.cm}\times \Big[
 \Gamma^{\Lambda}(1',2';3,4) \, \Gamma^{\Lambda}(3',4';1,2) 
 \nonumber \\
&& \mbox{} \hspace{-1.cm}-  \Gamma^{\Lambda}(1',4';1,3) \, \Gamma^{\Lambda}(3',2';4,2) 
 - (3 \leftrightarrow 4, 3' \leftrightarrow 4') \nonumber \\
&& \mbox{} \hspace{-1.cm}+  \Gamma^{\Lambda}(2',4';1,3) \, \Gamma^{\Lambda}(3',1';4,2)
 + (3  \leftrightarrow 4, 3'  \leftrightarrow 4') 
 \, \Big] \; .\label{flowGamma}
\end{eqnarray}
The labels $1$, $2$, etc., are a shorthand for the quantum numbers of the 
one-particle basis $j,\sigma$ and the Matsubara frequencies and 
the summation stands for a sum over the quantum numbers and an integral 
over the frequency. The full propagator ${\mathcal G}^{\Lambda}$ is given by
the Dyson equation
\begin{equation}
\label{Gdef}
 {\mathcal G}^{\Lambda} = \left[ 
( {\mathcal G}_0^{\Lambda})^{-1} - \Sigma^{\Lambda} \right]^{-1}
\end{equation}
and the so-called single-scale propagator $S^{\Lambda}$ by 
\begin{eqnarray}
\label{singlescale}
 S^{\Lambda}  =
  {\mathcal G}^{\Lambda} \left[ \frac{\partial}{\partial \Lambda} 
\left( {\mathcal G}_0^{\Lambda}\right)^{-1} \right] 
  {\mathcal G}^{\Lambda}  \; . 
\end{eqnarray}
The order of the projection onto the system of dots and the
introduction of a cutoff can equally be interchanged, leading to the same
set of flow equations.

We next implement our second approximation:
the frequency-dependent flow of the renormalized 2-particle vertex 
$\Gamma^{\Lambda}$ is replaced by its value at vanishing (external) 
frequencies, such that $\Gamma^{\Lambda}$ remains frequency
independent.
Since the bare interaction is frequency independent, neglecting the
frequency dependence leads to errors only at second order (in the 
interaction strength) for the self-energy, and at third order for
the vertex function at zero frequency.
As a consequence, also the self-energy becomes frequency independent.
Then  
$\varepsilon_{j,\sigma}+\Sigma_{j,\sigma;j,\sigma}$ can be viewed 
as the effective  (single-particle) level position and 
$t_{j,j'}+\Sigma_{j,\sigma;j',\sigma}$, with $j\neq j'$ 
as the effective  (single-particle) hopping between the levels. 
In general both depend on all parameters of the problem, in 
particular the two-particle interaction.
This 
interpretation will later be helpful to gain further insights in 
our results.

For a sharp frequency cutoff the frequency integrals on the right-hand 
side (rhs) of the flow equations (\ref{flowS}) and 
(\ref{flowGamma}) can be  carried out analytically. At this point 
one has to deal with products of delta functions  
$\delta(|\omega| - \Lambda)$ and expressions involving step 
functions $\Theta(|\omega| - \Lambda)$. These at first sight 
ambiguous expressions are well defined and unique if the sharp 
cutoff is implemented as a limit of increasingly sharp broadened 
cutoff functions $\Theta_{\epsilon}$, with the broadening  parameter 
$\epsilon$ tending to zero. The expressions can then be conveniently 
evaluated by using the following relation,\cite{Morris} 
valid for arbitrary continuous functions $f$:
\begin{equation}
 \delta_{\epsilon}(x-\Lambda) \, f[\Theta_{\epsilon}(x-\Lambda)] \to
 \delta(x-\Lambda) \int_0^1 f(t) \, dt \; ,
\end{equation}
where $\delta_{\epsilon} = \Theta'_{\epsilon}$. 
Note that the functional form of $\Theta_{\epsilon}$ for finite $\epsilon$ 
does not affect the result in the limit $\epsilon \to 0$.
For the approximate flow equations we then obtain
\begin{equation}
 \frac{\partial}{\partial\Lambda} \Sigma^{\Lambda}_{1',1} =
 - \frac{1}{2\pi} \sum_{\omega = \pm \Lambda} \sum_{2,2'} \,
 e^{i\omega 0^+} \, \tilde {\mathcal G}^{\Lambda}_{2,2'}(i\omega) \,
 \Gamma^{\Lambda}_{1',2';1,2} 
\label{finalflowsigma}
\end{equation}
and 
\begin{widetext}
\begin{eqnarray}
\frac{\partial}{\partial\Lambda} \Gamma^{\Lambda}_{1',2';1,2}  =  
\frac{1}{2\pi} \, 
 \sum_{\omega = \pm\Lambda} \, \sum_{3,3'} \, \sum_{4,4'}
 \Big\{ \frac{1}{2} \, \tilde {\mathcal G}^{\Lambda}_{3,3'}(i\omega) \, 
  \tilde {\mathcal G}^{\Lambda}_{4,4'}(-i\omega) && \!\!\!\!\!
 \Gamma^{\Lambda}_{1',2';3,4} \, \Gamma^{\Lambda}_{3',4';1,2} 
+ \tilde {\mathcal G}^{\Lambda}_{3,3'}(i\omega) \, \tilde
 {\mathcal G}^{\Lambda}_{4,4'}(i\omega)  \nonumber \\
&&  \times
 \left[ - \Gamma^{\Lambda}_{1',4';1,3} \, \Gamma^{\Lambda}_{3',2';4,2} 
        + \Gamma^{\Lambda}_{2',4';1,3} \, \Gamma^{\Lambda}_{3',1';4,2}
 \right] \Big\}
\label{finalflowgamma}
\end{eqnarray}
\end{widetext}
where the lower indexes $1$, $2$, etc.\  now stand for the 
single-particle quantum numbers $j,\sigma$ (not frequencies) and
\begin{equation}
\label{Glambdadef}
 \tilde  {\mathcal G}^{\Lambda}(i\omega) = 
 \left[  {\mathcal G}_0^{-1}(i\omega) - \Sigma^{\Lambda} \right]^{-1} \; .
\end{equation}
Note that in contrast to $ {\mathcal G}^\Lambda$ in Eq.~(\ref{Gdef}), $\tilde
 {\mathcal G}^{\Lambda}$ contains the cutoff-free noninteracting propagator. 
Thus, by taking a sharp cutoff the explicit cutoff function completely 
disappears from the flow equation, making it smooth and easy to integrate 
numerically.\cite{EnssThesis}
At the initial cutoff $\Lambda=\infty$ the flowing 2-particle vertex
$\Gamma^{\Lambda}_{1',2';1,2}$  is given by the antisymmetrized
interaction (see below). Equation~(\ref{flowGamma})  preserves 
the antisymmetry under the exchange of the first and second two 
indices. This reduces the number of independent 
$\Gamma^{\Lambda}_{1',2';1,2}$.
Furthermore, Eq.~(\ref{finalflowgamma}) preserves the spin
conservation of the Hamiltonian. One can either use these 
symmetries to reduce the number of both, flow equations and 
independent summations on the rhs of 
Eq.~(\ref{finalflowgamma}), or directly implement this equation for simplicity.
In complex cases the first will considerably speed up the 
numerical integration of the flow equations.   
To exemplify this we  consider the single-level case. Then the 
only physical interaction is the one between  spin up and spin 
down electrons and Eq.~(\ref{finalflowgamma}) reduces to a single 
equation with no sum over quantum numbers left on the rhs. 

The flow is determined uniquely by the differential flow equations 
and the initial conditions at $\Lambda = \infty$. 
The flow of the 2-particle 
vertex starts from the antisymmetrized bare 2-particle
interaction\cite{NegeleOrland}  $I_{1',2';1,2}$
while $m$-particle vertices of higher order vanish 
in the absence of bare $m$-body interactions with $m>2$.
The self-energy at $\Lambda = \infty$ is given by the bare one-particle 
potential, that is, by those one-particle terms which are not
included already in ${\mathcal G}_0$. For fixed quantum numbers 
$j,\sigma$ in Eq.~(\ref{dotham}) these are terms of the form 
$-\sum_{\sigma',j'} U_{j,j'}^{\sigma,\sigma'} /4$.  
In a numerical solution the flow starts at some large finite
initial cutoff $\Lambda_0$. Here one has to take into account that, 
due to the slow decay of the rhs of the flow equation
for $\Sigma^{\Lambda}$ at large $\Lambda$, the integration of the flow from 
$\Lambda = \infty$ to $\Lambda= \Lambda_0$ yields a contribution which does
not vanish in the limit $\Lambda_0 \to \infty$, but rather tends to a
finite constant.
Since $\tilde  {\mathcal G}^{\Lambda}_{2,2'}(i\omega) 
\to \delta_{2,2'}/(i\omega)$ for
$|\omega| = \Lambda \to \infty$, this constant is easily determined as
\begin{eqnarray}
 - \frac{1}{2\pi} \lim_{\Lambda_0 \to \infty} \int_{\infty}^{\Lambda_0}
 d\Lambda \sum_{\omega = \pm\Lambda} &&  \!\!\!\!\! 
\sum_{2,2'} e^{i\omega 0^+} \,
 \frac{\delta_{2,2'}}{i\omega} \, I_{1',2';1,2} \nonumber \\
&& = \frac{1}{2} \sum_{2} I_{1',2;1,2} \; .
\end{eqnarray}
This exactly cancels the bare one-particle term (see above). 
In summary, the initial conditions for the self-energy and the
2-particle vertex at $\Lambda = \Lambda_0 \to \infty$ are
\begin{eqnarray}
\label{inisigma}
 \Sigma^{\Lambda_0}_{1,1'} &=& 0 \\[2mm]
\label{inigamma}
 \Gamma^{\Lambda_0}_{1',2';1,2} &=& I_{1',2';1,2} \; .
\end{eqnarray}
For the flow at $\Lambda < \Lambda_0$ the factor $e^{i\omega 0^+}$ in 
Eq.~(\ref{finalflowsigma}) can be discarded.

In the following sections we use the above fRG-based approximation 
scheme to study the conductance, the transmission phase, 
and the level occupancies for the geometries sketched 
in Fig.~\ref{fig1}. For the 
single-dot case we also apply an even simpler scheme for comparison, where the flowing
2-particle vertex is replaced by its initial value. The system of flow 
equations then reduces to  Eq.~(\ref{finalflowsigma}). 

\section{Single dots}
\label{singledot}

As a first example we study the single dot case of Fig.~\ref{fig1} A. 
Since $N=1$ we suppress the dot index $j=1$. 
Furthermore, the Hamiltonian conserves the spin and the self-energy
and Green function is diagonal in the spin index $\sigma$.  
The same holds for the other setups studied. 
The level energies are 
$\varepsilon_{\uparrow}=V_g-{\mathcal H}/2$, 
$\varepsilon_{\downarrow}=V_g+{\mathcal H}/2$, 
where ${\mathcal H}$ denotes a magnetic field, and the 
interaction between up and down electrons occupying the dot 
level is $U$. We here consider the case with left-right symmetry of
the level-lead hoppings $t^L=t^R=t'$, but the results can easily 
be generalized. 
Note that due to the shift of the density by $-1/2$ in 
the last line of Eq.~(\ref{dotham}), $V_g=0$ corresponds to
half filling of the dot. 
The projected noninteracting propagator is given by\cite{Hewson}
\begin{eqnarray}
\label{singledotprop0}
  {\mathcal G}_{0,\sigma} (i
  \omega) = \frac{1}{i \omega -(V_g + \sigma {\mathcal H}/2) + 
i \Gamma \sgn(\omega)}  \; ,
\end{eqnarray}
where on the rhs $\sigma = \,\uparrow\, = +1$ and  
$\sigma = \,\downarrow\, = -1$. We have performed the wide-band limit that leads 
to an energy-independent hybridization $\Gamma=\Gamma^L+\Gamma^R=2 \pi t'^2 
\rho_{\text{lead}}$. This  is achieved 
by replacing $\tau \to  \eta \tau$ and 
$t' \to \sqrt{\eta} t'$ and taking $\eta \to \infty$. 
The flow equations for the effective level positions 
$V_\sigma^\Lambda=V_g+ \sigma {\mathcal H}/2+\Sigma_{\sigma}^\Lambda$
are   
\begin{eqnarray}
\label{sigmasingledot}
\frac{\partial }{\partial \Lambda}V_\sigma^\Lambda & = & - 
\frac{U^\Lambda}{2 \pi} \, \sum_{\omega=\pm \Lambda} 
\tilde {\mathcal G}^{\Lambda}_{\bar \sigma}(i
\omega) \nonumber \\
& = & 
\frac{U^\Lambda \, V_{\bar\sigma}^\Lambda/\pi}{(\Lambda+ \Gamma)^2
+(V_{\bar \sigma}^\Lambda)^2} \, ,
\end{eqnarray}   
with the initial condition $V_\sigma^{\Lambda=\infty}=V_g+\sigma {\mathcal H}/2$ 
and $\bar \sigma$ denoting the complement of $\sigma$. 
The cutoff-dependent propagator $\tilde  {\mathcal G}^{\Lambda}_{\sigma}(i
\omega) $ follows from $  {\mathcal G}_{0,\sigma} (i \omega)$ by replacing 
$V_g + \sigma {\mathcal H} /2 \to V_\sigma^\Lambda$.
As already mentioned above, using symmetries the flow of the 2-particle 
vertex can be reduced to a single equation for the effective interaction
between spin up and spin down electrons $U^\Lambda$. It is defined by 
$U^\Lambda = \Gamma^\Lambda_{\sigma,\bar \sigma;\sigma, \bar \sigma}$,
has the flow equation 
\begin{eqnarray}
\label{Usingledot}
\frac{\partial }{\partial \Lambda}U^\Lambda  
 & \! \! \! \! =  \!\! \!\!  &  \frac{(U^\Lambda)^2}{2 \pi} \!\!  
\sum_{\omega=\pm \Lambda} \! \! \left[ 
\tilde  {\mathcal G}^{\Lambda}_{\uparrow}(i \omega) \,
\tilde   {\mathcal G}^{\Lambda}_{\downarrow}(-i \omega)
+\tilde  {\mathcal G}^{\Lambda}_{\uparrow}(i \omega) \,
\tilde   {\mathcal G}^{\Lambda}_{\downarrow}(i \omega)
  \right]
\nonumber \\ 
& \!=  \! &  \frac{2\, \left(U^\Lambda\right)^2\, 
V_\uparrow^\Lambda \, V_\downarrow^\Lambda/\pi}{\left[ 
(\Lambda+ \Gamma)^2
+(V_{\uparrow}^\Lambda)^2 \right]\left[ 
(\Lambda+ \Gamma)^2
+(V_{\downarrow}^\Lambda)^2 \right]} \; ,
\end{eqnarray}   
and initial condition $U^{\Lambda=\infty} = U$. 

Within this approximation the dot spectral function at the end of the fRG 
flow is given by  
\begin{eqnarray}
\label{dotspecfu}
\rho_\sigma( \omega) = \frac{1}{\pi} \; 
\frac{\Gamma}{\left( \omega - V_\sigma \right)^2 + \Gamma^2} \; ,
\end{eqnarray}
with $V_\sigma = V^{\Lambda=0}_\sigma$, that is a Lorentzian of full 
width $2\Gamma$ and height $1/(\pi \Gamma)$ centered around $V_\sigma$.  
For the single dot, instead of using Eq.~(\ref{LB3}) the transmission can
be expressed in terms of the dot spectral weight at the chemical 
potential\cite{Meir}
\begin{eqnarray}
\label{Gfromspec}
G_\sigma(V_g) = \frac{e^2}{h} \; \pi \Gamma \; \rho_\sigma( 0) \; .
\end{eqnarray}  
Within our approximation the spectral weight and thus the conductance 
directly follows from the effective level position $V_\sigma $ 
at the end of the fRG flow.

In the following we consider strong couplings 
$U/\Gamma \gg 1$ and start out with the case ${\mathcal H}=0$. 
The latter implies that $V_\uparrow^\Lambda = V_\downarrow^\Lambda
=V^\Lambda$ and $G_\uparrow = G_\downarrow$.
The spectral function computed at $V_g=0$ using 
NRG (that is believed to be very close to the exact spectral function) shows a sharp 
Kondo resonance of height $1/(\pi \Gamma)$ and width $T_K$, where $T_K$
is the Kondo temperature, located around $\omega=0$. It has two additional 
broader features at $\omega=\pm U/2$ (Hubbard bands). For 
$-U/2 \lessapprox V_g \lessapprox U/2$ the Kondo peak is pinned 
at (close to) the chemical potential and has fixed height.\cite{Hewson} 
Using Eq.~(\ref{Gfromspec}) this leads to the plateau of width $U$ 
in $G(V_g)$ discussed in the introduction. Although our approximate 
spectral function neither 
shows the narrow Kondo resonance nor the Hubbard bands, we show 
that by a pinning of $V_\sigma$ it captures the pinning 
of the spectral weight. We reproduce the  line shape of the
conductance quantitatively up to very large $U/\Gamma$.   

\begin{figure}[tb]
\begin{center}
\includegraphics[width=0.48\textwidth,clip]{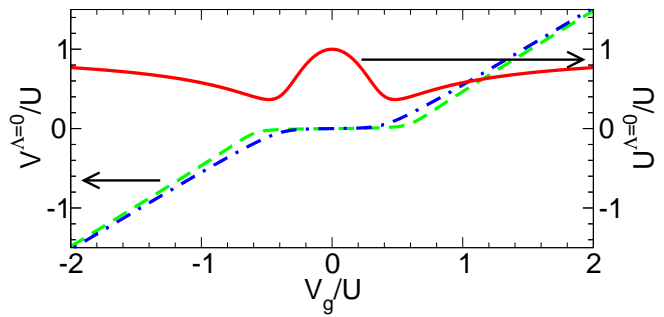}
\end{center}
\vspace{-0.6cm}
\caption[]{(Color online) 
Gate Voltage dependence of the effective level position and
  interaction for a single-level dot at 
$U/\Gamma=4 \pi$ and ${\mathcal H}=0$. Dashed line:
  $V=V^{\Lambda=0}$ without flow of the interaction. Dashed-dotted
  line: $V=V^{\Lambda=0}$ with flow of the interaction. Solid line: 
  effective interaction $U^{\Lambda=0}$.
\label{fig2}}
\end{figure}
\begin{figure}[tb]
\begin{center}
\includegraphics[width=0.45\textwidth,clip]{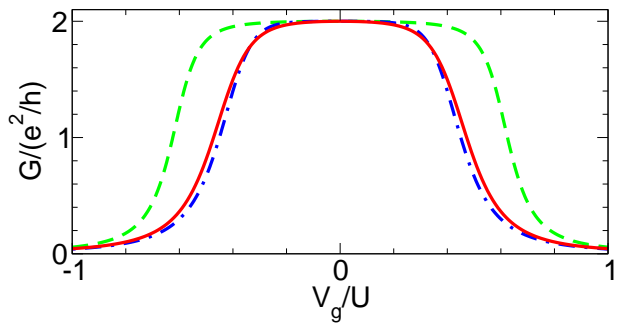}
\end{center}
\vspace{-0.6cm}
\caption[]{(Color online) 
Gate Voltage dependence of the linear conductance for the same
parameters as in Fig.~\ref{fig2}. Solid line: exact Bethe
  ansatz solution from Ref.~\onlinecite{Gerland}. Dashed line: fRG
  approximation without flow of the interaction. Dashed-dotted line:
  fRG approximation with flow of vertex.
\label{fig3}}
\end{figure}

To gain some analytical insight we first replace 
$U^\Lambda \to U$ and solve Eq.~(\ref{sigmasingledot}). 
The solution is obtained in implicit form 
\begin{eqnarray}
\label{sol}
\frac{v J_1(v)-\gamma J_0(v)}{v Y_1(v)-\gamma Y_0(v)} =
\frac{J_0(v_g)}{Y_0(v_g)} \; ,
\end{eqnarray} 
with $v= V \pi/U$, $v_g= V_g \pi/U$, $\gamma= 
\Gamma \pi/U$, and Bessel functions $J_n$, $Y_n$. 
For $|V_g| < V_c$, with $v_c=V_c \pi/U$ being the first zero of $J_0$, 
i.e., $V_c=0.7655 \;  U$, this equation has a solution with
a small $|V|$. For $U \gg \Gamma$ the crossover to a
solution with $|V|$ being of order $U$ (for $|V_g| > V_c$) 
is fairly sharp. Expanding both sides of Eq.~(\ref{sol}) for small
$|v|$ and $|v_g|$ gives $V = V_g \exp{[-U/(\pi \Gamma)]}$. Inserting 
this in Eq.~(\ref{dotspecfu}) leads to an exponential pinning of the 
spectral weight at $\mu$. In Fig.~\ref{fig2} we show $V(V_g)$ 
for $U/\Gamma = 4 \pi$ as the dashed line. For $|V_g| \gg U/2$
we find $V = V_g - \sgn(V_g) \, U/2$.
Applying Eq.~(\ref{Gfromspec}) the exponential pinning of the 
spectral weight at $\mu=0$ for small $|V_g|$ and the sharp 
crossover to a $V$ of order $U$ when $|V_g|> V_c$ leads 
to the plateau in $G(V_g)$ shown as the dashed line in  
Fig.~\ref{fig3}. 
For $U \gg \Gamma $ the width of the plateau  
is $2 V_c = 1.531 U$, which is larger than the width $U$ found
from the exact Bethe ansatz solution\cite{Gerland}  
(solid line in Fig.~\ref{fig3}) and with NRG.

This can systematically be improved including the flow of the
interaction Eq.~(\ref{Usingledot}). As the dashed-dotted and solid
lines in Fig.~\ref{fig2} we show the effective level position and the
effective interaction at the end of the fRG flow. The dashed-dotted 
line in Fig.~\ref{fig3} is the 
resulting $G(V_g)$. Away from $V_g=0$ the effective 
interaction first decreases and the plateau becomes narrower. 
The approximate conductance then agrees quantitatively with the 
exact Bethe ansatz result (solid line in Fig.~\ref{fig3}). For decreasing 
$U/\Gamma$ the agreement of fRG and Bethe ansatz data 
systematically improves and for $U/\Gamma
\lessapprox 6$ the two curves can barely be distinguished. The agreement 
becomes only slightly worse than in Fig.~\ref{fig3} for 
$U/\Gamma$ as large as 25, the largest $U/\Gamma$ for which Bethe 
ansatz data are available in the literature.\cite{Gerland} Also 
for more complex dot geometries
including the flow of the effective interaction considerably improves
the agreement with NRG and from now on we exclusively use this truncation
scheme.   

\begin{figure}[tb]
\begin{center}
\includegraphics[width=0.45\textwidth,clip]{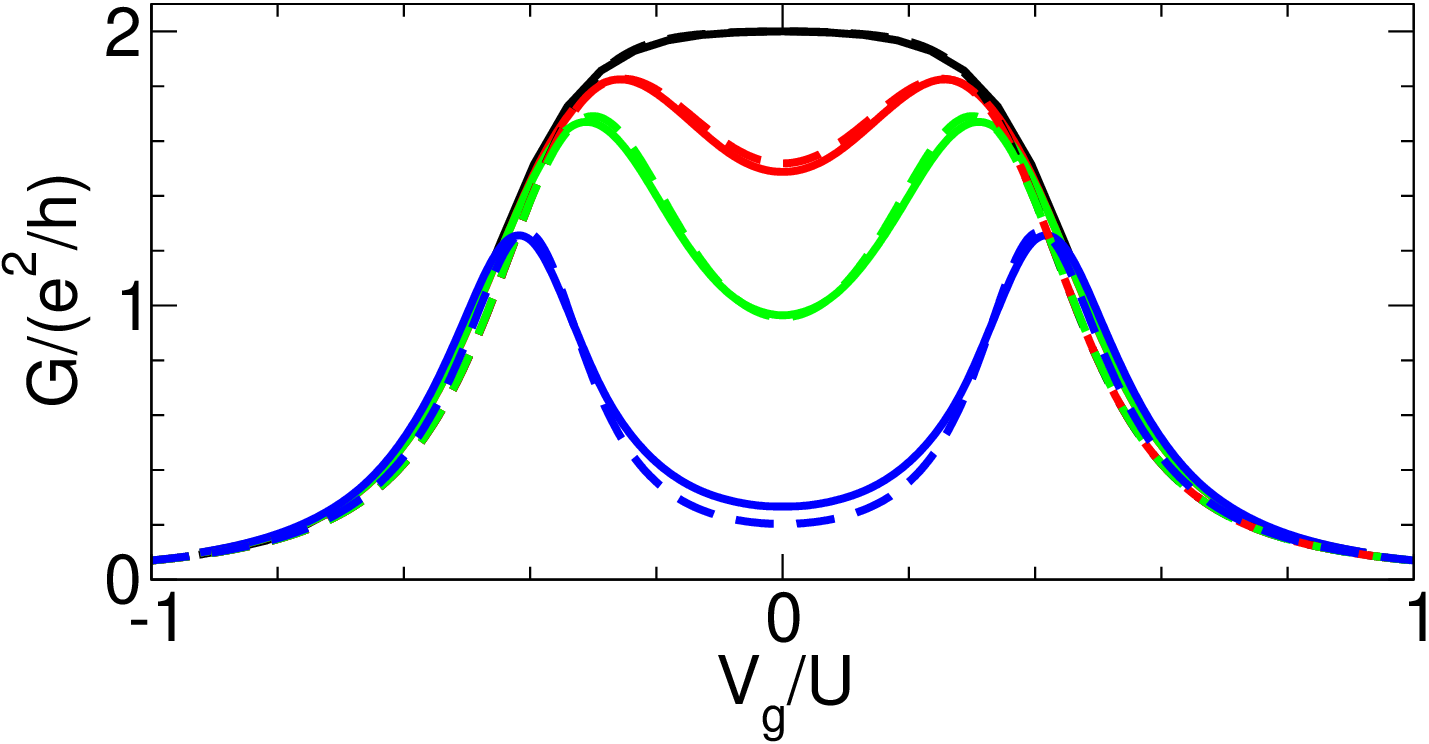}
\end{center}
\vspace{-0.6cm}
\caption[]{(Color online) 
Gate Voltage dependence of the total conductance $G$ of a
  single dot with $U/\Gamma=3\pi$ and ${\mathcal H}/\Gamma=0$, 
$0.058$, $0.116$, $0.58$ from top to bottom. In units of the $V_g=0$ 
  Kondo temperature $T_K^{\text NRG}/\Gamma=0.116$ these fields correspond 
  to ${\mathcal H}=0$, $0.5 T_K^{\text NRG}$,
  $T_K^{\text NRG}$, and $5T_K^{\text NRG}$. Solid line: NRG data 
  from Ref.~\onlinecite{Theo2}. Dashed line: fRG
  approximation with flow of vertex.
\label{fig4}}
\begin{center}
\includegraphics[width=0.45\textwidth,clip]{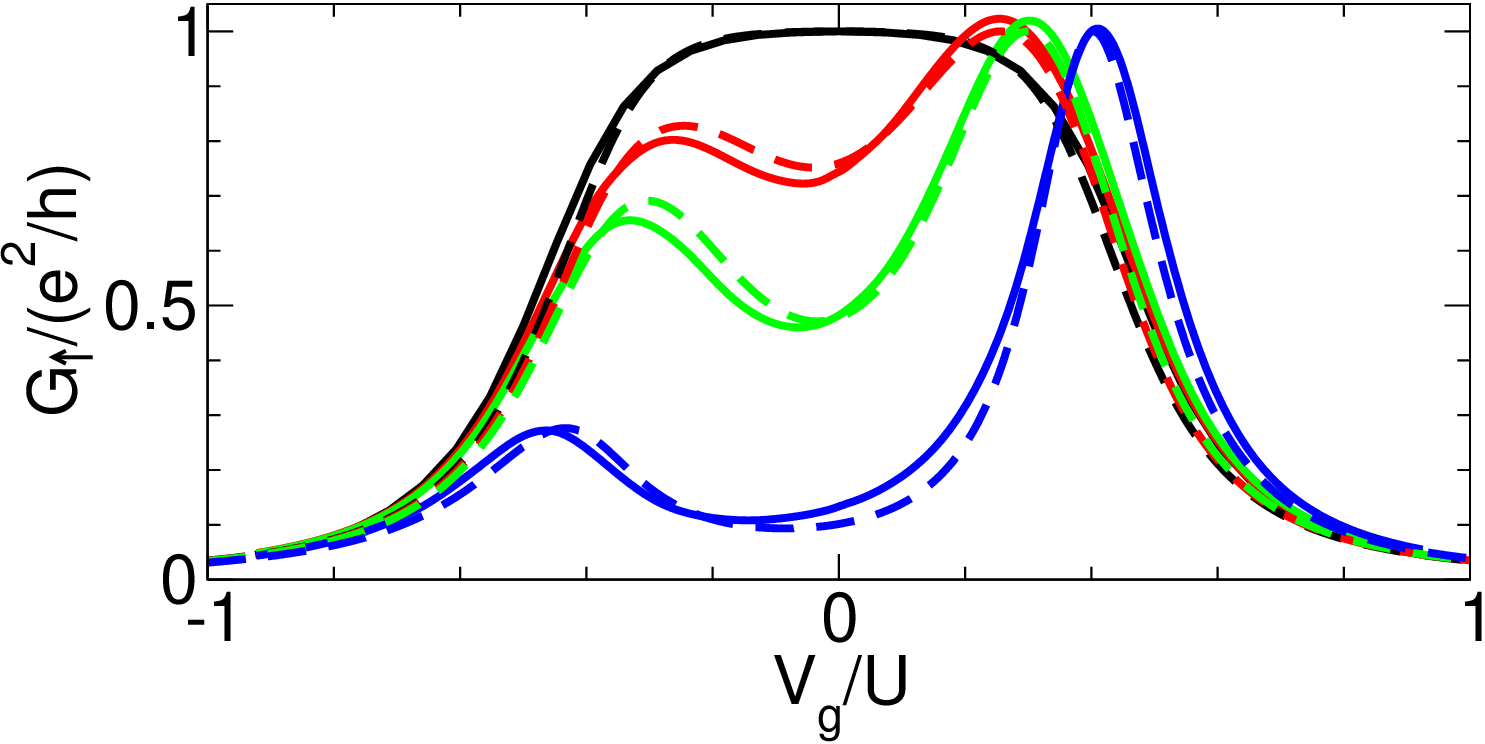}
\end{center}
\vspace{-0.6cm}
\caption[]{(Color online) 
Gate Voltage dependence of the partial conductance $G_\uparrow$ of a
  single dot for the same parameters as in  Fig.~\ref{fig4}.
\label{fig5}}
\end{figure}

We next consider the case of finite magnetic fields. For ${\mathcal H}>0$ the
Kondo resonance in the NRG solution of the spectral function splits in
two peaks with a dip at $\omega=0$, resulting in a dip of $G(V_g)$
at $V_g=0$. In Figs.~\ref{fig4} and \ref{fig5} we compare the total 
$G=G_\uparrow+G_\downarrow$ and partial $G_\uparrow$ conductance
obtained from our fRG truncation scheme including the flow of the
interaction and from NRG\cite{Theo2} for $U/\Gamma=3 \pi$ and
different ${\mathcal H}$. 
In the caption we give ${\mathcal H}$ in units of $T_K^{\text
  NRG}=0.116\Gamma$, where $T_K^{\text NRG}$
is determined from the width of the Kondo resonance at the
particle-hole symmetric point $V_g=0$ using NRG.\cite{Theo2} 
The agreement between NRG and fRG results is 
excellent. In particular, at $V_g=0$ the NRG and fRG data for 
${\mathcal H}=T_K^{\text NRG}$ are almost indistinguishable. 

\begin{figure}[tb]
\begin{center}
\includegraphics[width=0.45\textwidth,clip]{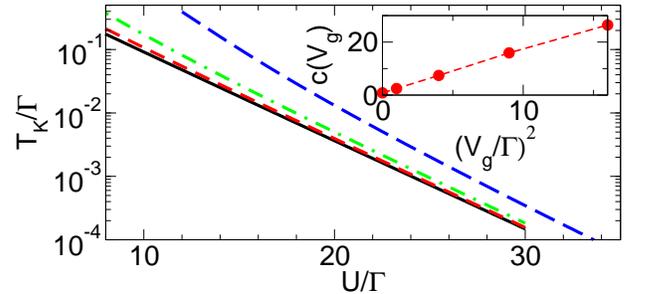}
\end{center}
\vspace{-0.6cm}
\caption[]{(Color online) 
The Kondo scale $T_K$ as a function of $U$ for different
$V_g$: $V_g=0$ (solid), $V_g/\Gamma=1$ (short dashed), $V_g/\Gamma=2$
(dashed-dotted), and $V_g/\Gamma=4$ (long dashed). {\it Inset:} The 
fitting parameter $c$ as a function of
$(V_g/\Gamma)^2$.
\label{fig5a}}
\end{figure}

This observation
suggest that our approximate scheme indeed contains the Kondo temperature
$T_K$ that depends exponentially on a combination of $U$ and the level
position\cite{Hewson} 
\begin{eqnarray}
\label{TKondo}
T_K^{\text exact} & \sim & \exp{\left[- \frac{\pi}{2 U \Gamma} 
\left|U^2/4-V_g^2\right|\right]}
\nonumber \\
 & = & \exp{\left[- \left| \frac{\pi}{8} \frac{U}{\Gamma} - 
  \frac{\pi}2 \frac{V_g^2}{\Gamma^2} \frac{\Gamma}{U} \right| \right] }
\; .
\end{eqnarray}
The prefactor of the exponential depends on the details of 
the model considered. To leading order its $U$ and $V_g$ dependence 
can be neglected. 
To investigate the appearance of an exponentially small energy scale
within our approximation we define a Kondo scale $T_K(U,V_g)$ by 
the magnetic field required to suppress the total conductance down
to half the unitary limit $G(U,V_g,{\mathcal H}=T_K)=e^2/h$. For fixed
$U\gg \Gamma$ this definition is meaningful for gate voltages which 
for ${\mathcal H}=0$ are in the conductance plateau. In  Fig.~\ref{fig5a}
we show $T_K(U,V_g)$ for different $V_g$ as a
function of $U$ on a linear-log scale. The curves can be
fitted by a function of the form [see Eq.~(\ref{TKondo})] 
\begin{eqnarray*}
f(U/\Gamma)=a \exp{\left[- \left| b \frac{U}{\Gamma} - c
      \frac{\Gamma}{U} \right| \right]} 
\end{eqnarray*}
with $V_g$ dependent coefficients $a$, $b$, and $c$. On the scale of
Fig.~\ref{fig5a} the original data and the fits are
indistinguishable. We find that $b(V_g)$ barely changes with $V_g$
and is given by $b \approx 0.32$, which is close to the exact value
$\pi/8 \approx 0.39$ [see Eq.~(\ref{TKondo})]. Furthermore, the prefactor 
$a$ depends only weakly on $V_g$ and $c(V_g)$ increases approximately
quadratically with $V_g/\Gamma$ as shown in the inset of
Fig.~\ref{fig5a}. 
Both these results are consistent with the behavior of the exact 
Kondo temperature  Eq.~(\ref{TKondo}). We thus conclude that 
$T_K(U,V_g)$ can be estimated from the 
${\mathcal H}$ dependence of $G$ obtained within the fRG. The Kondo
temperature can also be obtained from the local spin
susceptibility.\cite{Hewson} Computing this using our fRG scheme leads
to results equivalent to the ones discussed above.  
 
For the single dot at $T=0$ the exact conductance, transmission
phase, and dot occupancies are directly related by a 
generalized Friedel sum rule:\cite{Hewson}
$G_\sigma/(e^2/h)=\sin^2{(\pi \left< n_\sigma \right> )}$,
$\alpha_\sigma=\pi \left< n_\sigma \right>$. 
As $0 \leq \left< n_\sigma \right> \leq 1$ the argument of $\sin^2$ is
restricted to a single period and the relation between $G_\sigma$, $\left<
  n_\sigma \right>$, and $\alpha_\sigma$ is unique. 
In many approximation schemes the Friedel sum rule does not hold
exactly. Within our method we map the many-body problem onto an 
effective single-particle one for which the Friedel sum rule is
fulfilled.  
For gate voltages within the ${\mathcal H}=0$ conductance plateau the
(spin independent) dot filling is $1/2$ and the (spin independent) 
phase is $\pi/2$. For sufficiently large $U/\Gamma$ the crossover to 
$\left< n_\sigma \right>=1$ and $\alpha_\sigma=\pi$ to the left of the
plateau as well as  $\left< n_\sigma \right>=0$ and $\alpha_\sigma=0$ to the
right is fairly sharp.

As a more complex single-dot problem we study the case of a dot
embedded between two leads, that are also coupled directly (see Fig.~\ref{fig1}
C). In that case the conductance shows the characteristics 
of both the Kondo and the Fano effect, as was discussed earlier 
based on NRG results.\cite{Hofstetterkondofano}
We here focus on ${\mathcal H}=0$. 

To derive expressions for the dot Green function and the conductance
we first consider a three-site system of the dot and the last sites of
the left and right leads. The effect of the other lead sites is taken
into account by projecting them out, which leads to a single-particle
potential on the last sites of the leads similar to Eq.~(\ref{leadpotdef}) 
but with $t_j^l$ replaced by $\tau$. At $U=0$ the projected
``effective Hamiltonian'' [see Eq.~(\ref{projection})] for both spin 
directions reads (for the hopping between the leads and dot we 
still assume $t^L = t^R =t'$)
\begin{eqnarray}
\label{effhamkondofano}
h_\sigma(i\omega) = \left( \begin{array}{ccc} 
 \tau^2 \text{g}_{\text{lead}}(i\omega) & t_{LR} &  - t' \\
t_{LR} & \tau^2 \text{g}_{\text{lead}}(i\omega) & - t' \\
-t' & -t' & V_g
  \end{array} \right)
\end{eqnarray}
in a basis left lead---right lead---dot.
We here consider $t',t_{LR} \in \mathbb R$ (for the model with an enclosed
magnetic flux, see below). The Green function matrix element entering 
Eq.~(\ref{LB3}) is given by the $1-2$ matrix element of the resolvent
of $h_\sigma(i\omega)$, Eq.~(\ref{effhamkondofano}). Within our truncation 
scheme the same holds for $U>0$ if $V_g$ is replaced by the effective 
dot level position $V^{\Lambda=0}$ in $h_{3,3}(i\omega)$. 
To determine the flow equations for
$V^\Lambda$ and $U^\Lambda$ we employ Eq.~(\ref{projection}) a second 
time and project onto the dot site. After taking the wide-band limit 
the cutoff-dependent dot Green function is
\begin{eqnarray}
\label{Greenkondofano}
 \tilde {\mathcal G}^\Lambda_{\sigma} (i \omega) = 
\left[i \omega - V^\Lambda + \Gamma \; 
\frac{\tilde t_{LR} + i \sgn(\omega)}{1+\tilde t_{LR}^2} 
\right]^{-1}  \; ,
\end{eqnarray} 
with $\tilde t_{LR} = \pi t_{LR} \rho_{\text{lead}}$. The first lines of 
Eqs.~(\ref{sigmasingledot}) and (\ref{Usingledot}) also hold in 
the presence of the direct hopping $t_{LR}$, which completes the
derivation of the flow equations. 

\begin{figure}[tb]
\begin{center}
\includegraphics[width=0.45\textwidth,clip]{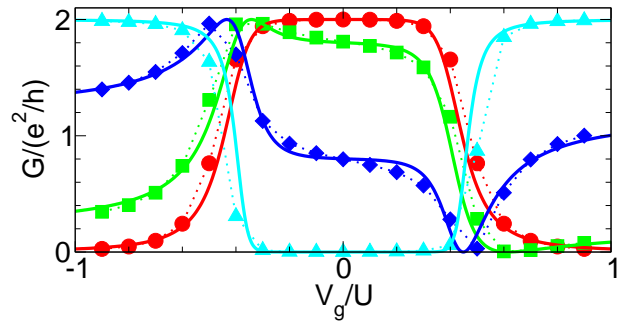}
\end{center}
\vspace{-0.6cm}
\caption[]{(Color online) 
Gate Voltage dependence of the total conductance $G$ of a
  single dot in the Fano geometry of Fig.~\ref{fig1} C with $U/\Gamma \approx
  5.05 \pi$, ${\mathcal H}=0$, and different ${\mathcal T}_{LR}$. Symbols: NRG data 
  from Ref.~\onlinecite{Hofstetterkondofano} (circles ${\mathcal T}_{LR}=0$,
  squares ${\mathcal T}_{LR} = 0.1$, diamonds 
  ${\mathcal T}_{LR} = 0.6$, triangles
  ${\mathcal T}_{LR}=1$). Solid lines: fRG approximation with flow of vertex.
\label{fig6}}
\end{figure}

At $U=0$ the interference of the direct energy independent transmission
channel and the resonant channel leads to a Fano line shape of
$G(V_g)$.\cite{Fano} 
The line shape resulting from the presence of both the Kondo
and the Fano effect is shown in Fig.~\ref{fig6}
for\cite{Hofstetterkondofano} $U/\Gamma \approx
5.05 \pi $ and different strength of the direct hopping between the
left and right lead. The latter is measured by the $U=0$ transmission 
probability ${\mathcal T}_{LR}(t_{LR})$ that would result if only the 
direct link would be present, that is for $t'=0$. 
For ${\mathcal T}_{LR} = 0$ one recovers the single-dot problem discussed
above. In the opposite limit of perfect direct transmission 
${\mathcal T}_{LR} =1$ the $U=0$ Fano anti-resonance at $V_g=0$ (see
the $q=0$ curve in Fig.~1 of Ref.~\onlinecite{Fano}) is extended to  
a broad region with $G(V_g) \approx 0$ due to the presence of the
Kondo effect. The results for $0 < {\mathcal T}_{LR} <1$ 
lie in between these two limiting cases. The fRG data again 
agree  well with the NRG results. 

Similar to the single dot without a direct link for the present geometry 
one can derive unique relations between the conductance, the 
transmission phase, and the occupancy.\cite{Hofstetterkondofano} 
Thus the last two observables do not carry any additional information
and we do not show them here.
In addition to the gate voltage  dependence also the dependence on 
an enclosed magnetic flux was studied. Replacing $t_{LR} \to \exp{(i
  \phi)} \; t_{LR}$ a flux can easily be included in our approach 
and we reproduce the results of Ref.~\onlinecite{Hofstetterkondofano}. 

Our results for the single-level dot show that although the fRG 
truncation scheme is  set up for small $U$, it works quantitatively 
for very large $U$. For ${\mathcal H}=0$ the method captures 
the aspect of Kondo physics essential to obtain the plateau 
in $G(V_g)$, namely the pinning of spectral weight. From this one 
can expect that the method gives reliable results also for more 
complex dot systems, as will be explored below. Using a truncation 
scheme in which the full frequency dependence of the 2-particle 
vertex is kept (leading to a frequency dependent self-energy) it 
was shown that one can also reproduce the Kondo resonance and 
Hubbard bands of the spectral function,\cite{Ralf} although with 
a much higher numerical effort.\cite{footnotekatanin}   

\section{Linear chains of dots}
\label{lindot}

We next study a linear chain of $N$ single-level dots as shown in Fig.~\ref{fig1}
B. For this problem one already has to specify a variety of parameters. 
The inter-dot hopping matrix elements might not only be restricted to 
nearest neighbors but extend over a longer range. Depending on the 
experimental realization of the chain in addition the 
hoppings of equal range might not all be the same but explicitly
depend on the pair of dots considered. Besides the on-site
interaction (that might depend on $j$) one could introduce 
longer-range interactions. The on-site energies $\varepsilon_{j,\sigma}$
might vary from site to site. Within our approach all these situations 
can be investigated, as the parameters only enter the projected 
noninteracting propagator and the initial conditions for the 
2-particle vertex.  We here consider the cases $N=3$ and $N=4$, but also  
$N=10$ has been studied with our method. We focus on ${\mathcal H}=0$,
nearest-neighbor inter-dot hoppings of equal
amplitude, equal on-site energies, and equal local interactions. We verified
that including a nearest-neighbor interaction and a small variation of
the inter-dot hoppings only weakly affects our results.  

At $U=0$ the projected ``effective Hamiltonian'' is a tridiagonal 
$N \times N$ matrix with off-diagonal entries 
\begin{eqnarray}
\label{chainoff}
h_{\sigma;j,j \pm 1}(i \omega)=-t  
\end{eqnarray}
and diagonal elements
\begin{eqnarray}
\label{chainon1}
h_{\sigma;j,j}(i \omega) = V_g - i \sgn(\omega) 
 \left( \Gamma_1^L \delta_{j,1} + \Gamma_N^R \delta_{j,N}
 \right) \; .
\end{eqnarray}
The (spin-independent) projected noninteracting propagator then 
follows from the matrix inversion 
\begin{eqnarray}
\label{matrixinversion}
 {\mathcal G}_{0,\sigma} (i\omega) = \left[i \omega- h_{\sigma}(i
  \omega)\right]^{-1} \;. 
\end{eqnarray}
The flow equations for the present setup are obtained by inserting
this $ {\mathcal G}_{0}$ into Eq.~(\ref{Glambdadef}) and using the resulting 
cutoff-dependent full propagator in Eqs.~(\ref{finalflowsigma})
and (\ref{finalflowgamma}). The initial conditions are given by 
Eqs.~(\ref{inisigma}) and (\ref{inigamma}).  
Although we start out with a purely local interaction, during the fRG flow
Eq.~(\ref{finalflowgamma}) all types of two-particle interactions 
not forbidden by spin conservation and antisymmetry of the vertex, 
e.g., long-range density-density interactions and pair hoppings,
are generated in
$\Gamma^{\Lambda}_{j_1',\sigma_1',j_2',\sigma_2';j_1,\sigma_1,j_2,\sigma_2}$.
We observe the same for the setups of Figs.~1 D and E. 
With this 2-particle vertex the flowing self-energy
Eq.~(\ref{finalflowsigma}) becomes a full $N\times N$ matrix. During the fRG flow 
the interaction does not only lead to a renormalization of the level 
positions and the nearest-neighbor hoppings, but also generates longer-range 
single-particle hoppings. Applying 
Eqs.~(\ref{LB1})-(\ref{projection}) the
conductance can be computed from the $1-N$ matrix element of the 
projected Green function at $\Lambda=0$ by (note that $\tilde
 {\mathcal G}^{\Lambda=0} = {\mathcal G}^{\Lambda=0}$)
 \begin{eqnarray}
\label{conductchain}
G_\sigma(V_g) = \frac{e^2}{h} \, 4 \Gamma_1^L 
\Gamma_N^R \left|  {\mathcal G}_{\sigma;1,N}^{\Lambda=0}(0) \right|^2  \; .
\end{eqnarray}
The occupancy of dot $j$ follows from integrating 
$ {\mathcal G}_{j,j}^{\Lambda=0}(i\omega)$ over $\omega$. 

\begin{figure}[tb]
\begin{center}
\includegraphics[width=0.45\textwidth,clip]{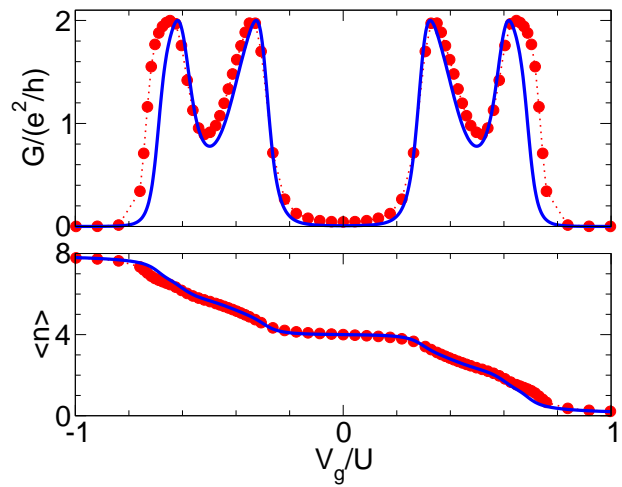}
\end{center}
\vspace{-0.6cm}
\caption[]{(Color online) 
Gate Voltage dependence of the conductance $G$ and the total
occupancy $\left< n \right>$ of a
  linear chain of $N=4$ spin-degenerate 
dots (see Fig.~\ref{fig1} B) with $U/\Gamma=2 \pi$ and
  $t/\Gamma  = 1$. Circles: NRG data from Ref.~\onlinecite{Oguri3}. 
  Solid line: fRG approximation with flow of vertex. 
\label{fig7}} 
\end{figure}

\begin{figure}[tb]
\begin{center}
\includegraphics[width=0.45\textwidth,clip]{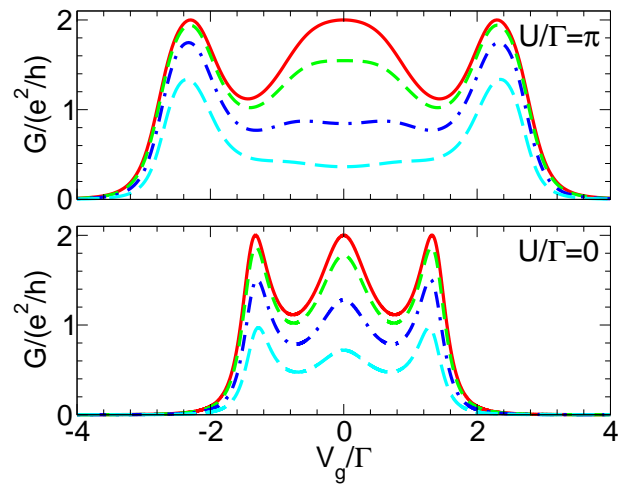}
\end{center}
\vspace{-0.6cm}
\caption[]{(Color online) 
Gate Voltage dependence of the conductance $G$ of a linear 
chain of $N=3$ spin-degenerate dots  
with asymmetric coupling to the
leads. The parameters are $U/\Gamma=\pi$ (upper panel), $U=0$
(lower panel), $t/\Gamma=1$, $\Gamma_1^L/\Gamma_3^R=1$ (solid line), 
$\Gamma_1^L/\Gamma_3^R=2$ (short dashed line),
$\Gamma_1^L/\Gamma_3^R=4$ (dashed-dotted line), and
$\Gamma_1^L/\Gamma_3^R=9$ (long dashed line). Note that here $V_g$ is
shown in units of $\Gamma$. 
\label{fig7a}} 
\end{figure}

For the special case of equal nearest-neighbor inter-dot hoppings $t$, 
equal hybridizations $\Gamma_1^L = \pi |t_1^L|^2 \rho_{\text{lead}} = \Gamma_N^R = \pi
|t_N^R|^2 \rho_{\text{lead}}=\Gamma/2$, equal local interactions $U$, and equal
$\varepsilon_{j,\sigma}=V_g$, $G(V_g)$ at $T=0$ was computed using
NRG.\cite{Oguri2,Oguri3} 
For this class of parameters the Hamiltonian has a high symmetry  and 
the conductance is related to scattering phase shifts. The latter   
can be extracted from energies of the system which allowed to obtain
results for a chain of up to four dots. In Fig.~\ref{fig7} we show a
comparison of the NRG and fRG results (for $G$ and the total dot
occupancy $\left< n \right> = \sum_{j,\sigma} \left< n_{j,\sigma}
  \right>$) for $N=4$,  
$U/\Gamma= 2 \pi$, and $t/\Gamma=1$. For gate voltages at which 
the four-site chain is occupied by an odd number of electrons 
one finds a transmission peak of unitary height. 
Each corresponds to transport through mainly one of the four dot
levels. For $U/\Gamma= 2 \pi$  the plateau-like 
line shape of the individual resonances induced  by the Kondo 
effect is only weakly developed. It becomes more pronounced if 
larger $U/\Gamma$ are considered. The overlap of the resonances
decreases for increasing $t/\Gamma$. 
For even $N$ the width 
of all the resonances is almost equal and of order $U/N$. In a wide
region around $V_g=0$, $G(V_g)$ is small because close to
half-filling of the chain charge fluctuations are strongly
suppressed. For odd $N$, as in Fig.~\ref{fig7a}, 
half-filling corresponds to an odd number of electrons on the chain. 
This implies a wide plateau of unitary height around $V_g$ due to the 
suppressed charge fluctuations (compare the solid lines in the upper
$U/\Gamma=\pi$ and lower panel $U=0$ of Fig.~\ref{fig7a}) and only 
the resonances away from $V_g=0$ have width $U/N$.\cite{Oguri2} 
The (spin-independent) transmission phase $\alpha_\sigma(V_g)$ and the
total occupancy are related by $\left< n \right> = 2
\alpha_\sigma/\pi$.\cite{Oguri3,footnotephases} 
Across every transmission resonance 
the phase continuously changes by $\pi$ in analogy to the behavior of
the phase at $U=0$. Note that in contrast to the
single-dot case $\alpha_\sigma$ is not restricted to the interval 
$[0,\pi]$.    

For $U/\Gamma$ smaller than the one of Fig.~\ref{fig7}
the agreement between  NRG and fRG further increases, e.g.~for 
$N=4$, $U/\Gamma=1.67 \pi$, and $t/\Gamma=4.167$ both data sets 
are almost indistinguishable (for the NRG data see Fig.~5 of
Ref.~\onlinecite{Oguri3}). We encounter serious deviations of 
the fRG data and NRG results if $U/\Gamma$ is significantly 
increased (i.e.~for $U/\Gamma=8.33$; for NRG data
see Fig.~4 of Ref.~\onlinecite{Oguri3}). In that 
case regions of gate voltages appear in which some  of the components of
the effective interaction become orders of magnitude larger than the
initial $U$ leading to a conductance that is too small. For these
$V_g$ it is no longer justified to neglect the frequency
dependence of the 2-particle vertex and the higher-order vertices. In
particular, this happens for $V_g$ on-resonance and in
addition for even $N$ in the deep valley around $V_g=0$. 
Indications of this breakdown of our approximation scheme for $U/\Gamma \gtrapprox  10$ 
can  already be observed in the $V_g=0$ valley, and the large 
$|V_g|$ resonances in Fig.~\ref{fig7}: in these gate-voltage regimes
our approximation significantly underestimates $G$.   

In Fig.~\ref{fig7a} we show how an increasing left-right asymmetry of
the $\Gamma_1^L$ and $\Gamma_N^R$ affects the conductance for $N=3$
for small to intermediate $t/\Gamma$.  
The parameters are $U/\Gamma=\pi$, with fixed $\Gamma=\Gamma_1^L + 
\Gamma_3^R$, and $t/\Gamma=1$. For comparison in the lower panel we
show the $U=0$ case. With increasing asymmetry the height of the peaks
decrease. While for fixed asymmetry the interaction leads to an
enhancement of the outer two peaks, it suppresses the peak located
around $V_g=0$. Therefore, at $U>0$ the central peak disappears quickly
if $\Gamma_1^L/\Gamma_3^R$ increases. It is important to note that
with increasing asymmetry the gate voltage dependence of the total
occupation $\left< n \right>$ and the transmission phase
$\alpha_\sigma$ (which are still related by $\left< n \right> = 2
\alpha_\sigma/\pi$) barely change. Thus these observables do not show
any indication of the disappearance of the central resonance.  
The same holds for the (effective) dot level-positions of the chain 
obtained  by diagonalizing the 
``effective Hamiltonian'' at the end of the fRG flow (computed in the
presence of the leads) after disconnecting the three sites from the 
leads. In particular, for all $\Gamma_1^L/\Gamma_3^R$ studied one of 
the level energies (eigenenergies of the isolated chain at 
$\Lambda=0$) goes through zero at $V_g=0$. For levels which 
are not too strongly (sufficient separation of the resonances) and not 
too asymmetrically coupled to the leads one  would thus 
expect a peak at $V_g=0$ (see e.g.~the $U=0$ case).
Surprisingly, during the fRG flow  the effective asymmetry of the
coupling to the outer
two levels (as obtained from the eigenvectors of the isolated chain at
$\Lambda=0$) is reduced compared to the initial value, while the
asymmetry of the central level is significantly increased by the
presence of the interaction. Furthermore, the sum of the effective
couplings of the central level to the left and right leads is
increased which implies that the level is broadened. The combined 
effect of the asymmetry and broadening explains the vanishing of 
the central peak. For increasing peak separation, that is for 
increasing $t/\Gamma$, the central resonance becomes more robust
against a left-right asymmetry.   

The above observation might be 
important for the interpretation of future experiments on linear 
chains of quantum dots because asymmetric dot-lead couplings 
are generic in  realistic dot systems. As shown, such an asymmetry 
might lead to a two peak structure although a chain of three sites 
is studied.

\section{Side-coupled double dots}
\label{sidecoupleddoubledot}

The side-coupled double dot of Fig.~\ref{fig1} D shows a very interesting
low-energy physics. Close to half-filling it is either dominated by 
a two-stage Kondo effect for small 
$t_{1,2}$ or the formation of a local ``molecular'' 
spin-singlet between the electrons on the two dots for large $t_{1,2}$. 
Both regimes can be distinguished by the temperature 
dependence of the $G(V_g)$ curves as is nicely discussed in 
Ref.~\onlinecite{Cornaglia} (see also Ref.~\onlinecite{Zitko}) 
based on NRG data. In the two-stage Kondo effect, 
first at high energies the spin of the embedded dot 
gets screened at the Kondo scale $T_K^0$ of the system with 
$t_{1,2}=0$. In the second stage the heavy quasi-particles screen 
the spin on the side-coupled dot at a much lower scale $T_0$.

The $U=0$ projected ``effective Hamiltonian'' for spin direction $\sigma$ 
reads (in the wide-band limit)
\begin{eqnarray}
\label{effhamsidecoupled}
h_\sigma(i\omega) = \left( \begin{array}{cc} 
\varepsilon_{1,\sigma} - i (\Gamma_1^L + \Gamma_1^R) 
\sgn(\omega)&  - t_{1,2} \\
- t_{1,2} & \varepsilon_{2,\sigma} 
  \end{array} \right) \; .
\end{eqnarray}
As for the linear chain of dots the projected noninteracting
propagator $ {\mathcal G}_0(i\omega)$ entering the flow equations
(\ref{finalflowsigma}) and (\ref{finalflowgamma}) follows by the
matrix inversion Eq.~(\ref{matrixinversion}). 
Starting out with a local interaction $U$ (and maybe a nearest-neighbor
density-density interaction $U'$) all two-particle interactions 
not excluded by
symmetries are generated (see above). Also in the self-energy matrix 
all four matrix elements flow, that is, both the level positions and 
the inter-dot hopping are renormalized. 
As for the single dot we can use Eq.~(\ref{Gfromspec}) to compute the
conductance from the spectral weight of the embedded dot with
$j=1$ taken at $\Lambda=0$.\cite{Cornaglia}  

\begin{figure}[tb]
\begin{center}
\includegraphics[width=0.45\textwidth,clip]{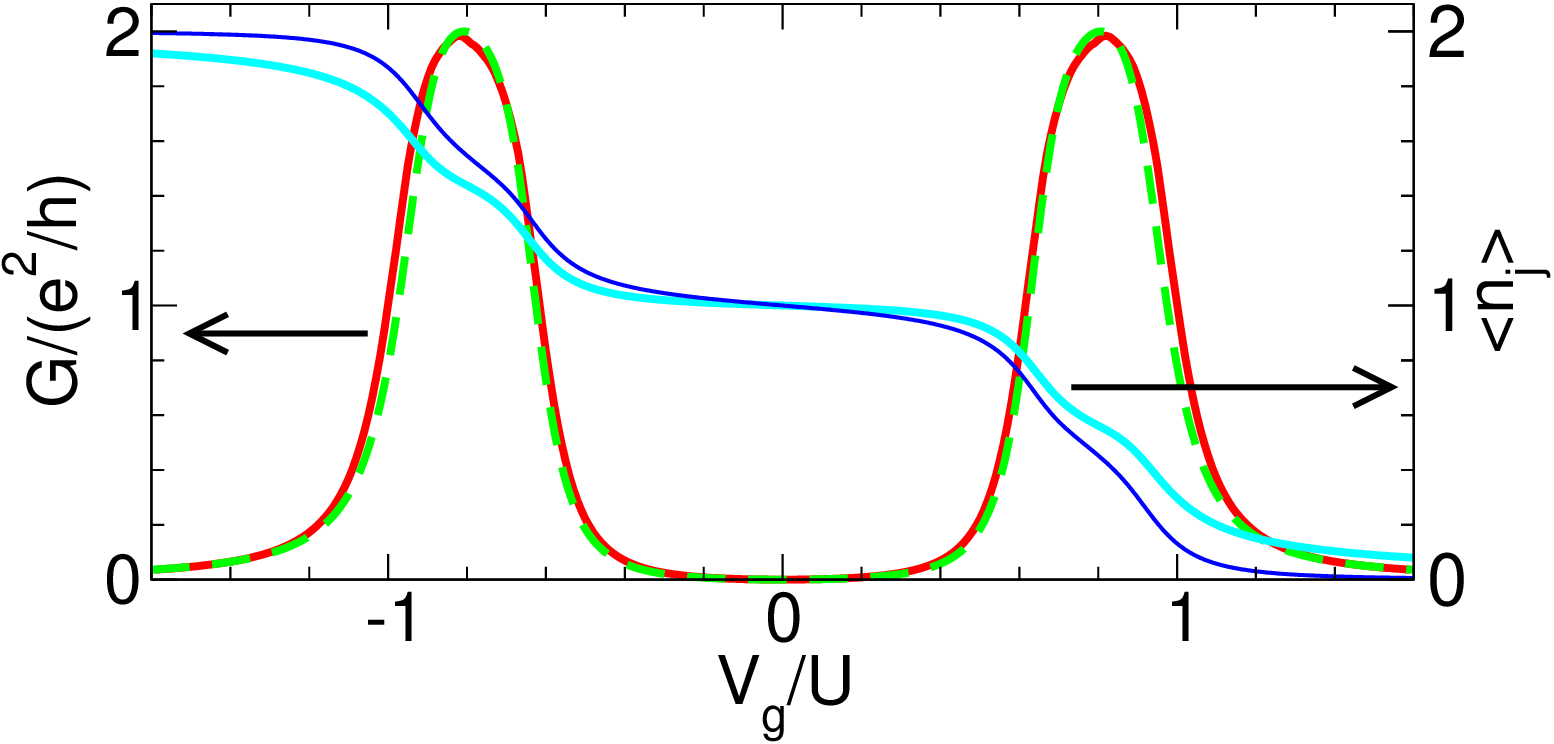}
\end{center}
\vspace{-0.6cm}
\caption[]{(Color online) 
Gate Voltage dependence of the total conductance $G$ and total 
occupancies $\left< n_{j} \right>$ of a spin-degenerate 
side-coupled double-dot system 
(see Fig.~\ref{fig1} D) with $U/\Gamma=8$ and $t_{1,2}/\Gamma=4$. 
Solid line with left y-axis: NRG data for $G$ 
  from Ref.~\onlinecite{Cornaglia}. Dashed line  with left y-axis: fRG
  approximation of $G$ with flow of vertex. Thick solid line with 
right y-axis: fRG approximation of $\left< n_1 \right>$ (embedded dot). 
Thin solid line with 
right y-axis: fRG approximation of $\left< n_2 \right>$ (attached dot). 
\label{fig8}}
\begin{center}
\includegraphics[width=0.45\textwidth,clip]{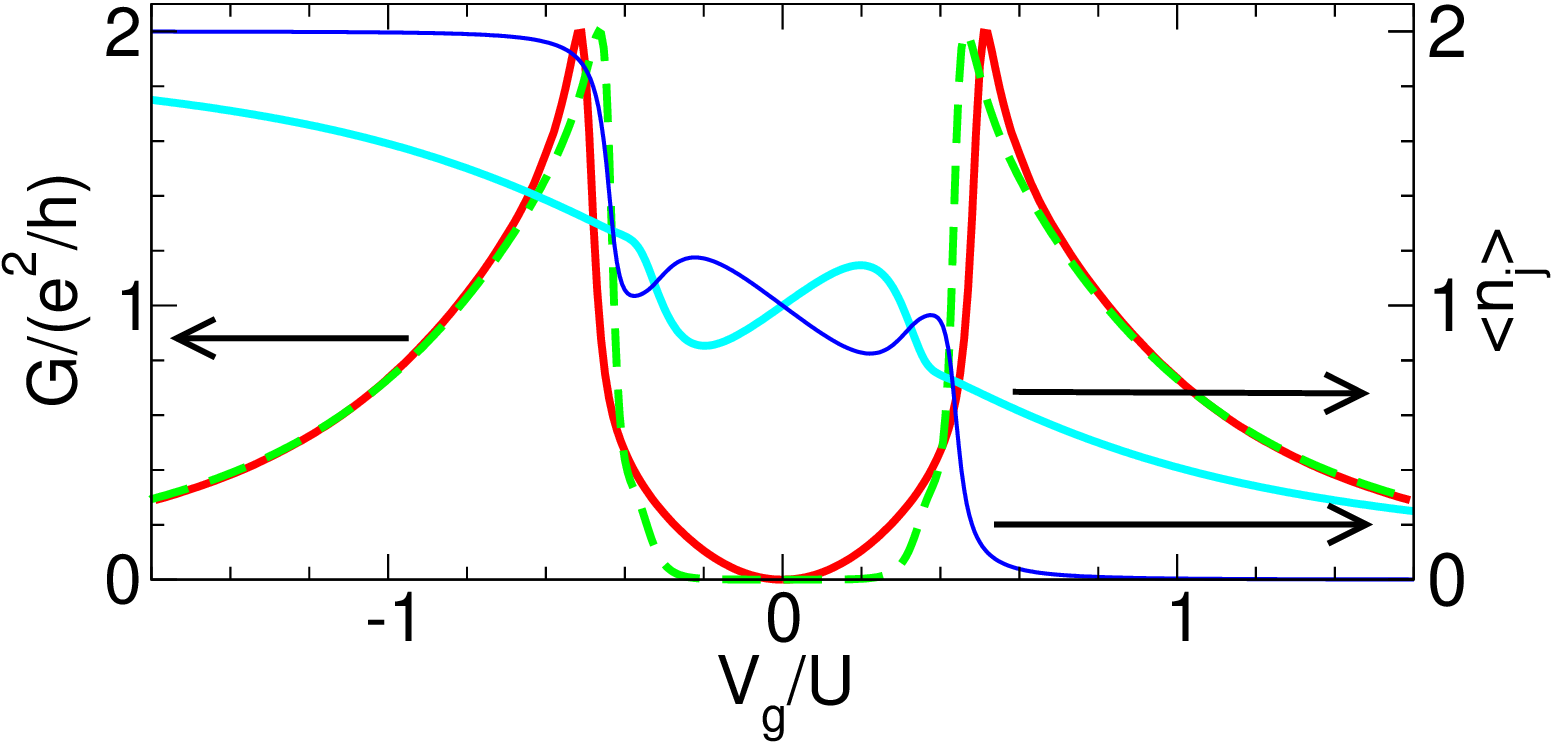}
\end{center}
\vspace{-0.6cm}
\caption[]{(Color online) 
As in Fig.~\ref{fig8} but for $U/\Gamma=2$ and $t_{1,2}/\Gamma=0.2$.
\label{fig9}}
\begin{center}
\includegraphics[width=0.45\textwidth,clip]{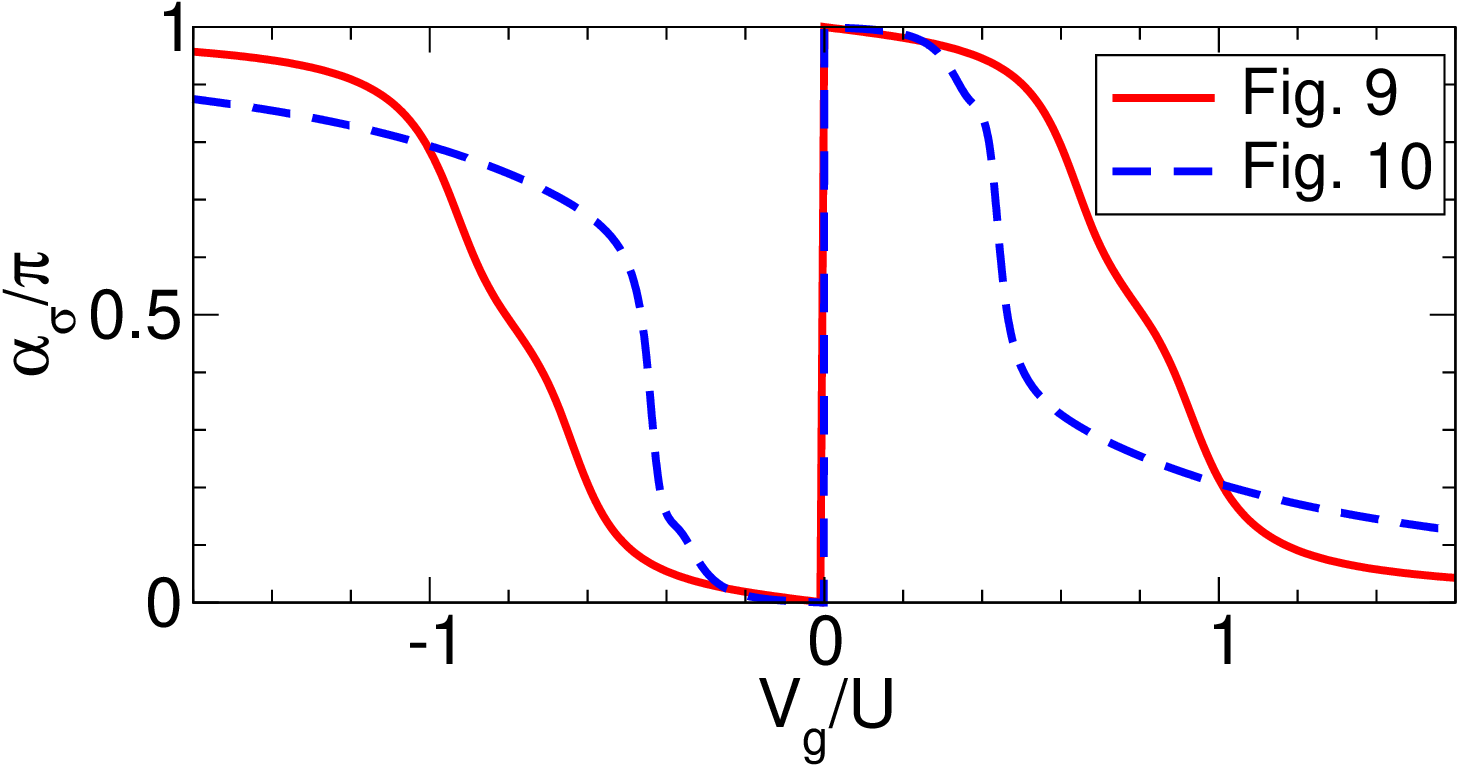}
\end{center}
\vspace{-0.6cm}
\caption[]{(Color online) 
Gate Voltage dependence of the (spin-independent) transmission phase
$\alpha_\sigma$ for the two parameter sets of Figs.~\ref{fig8} and
\ref{fig9}. 
\label{fig10}}
\end{figure}

In Figs.~\ref{fig8} and \ref{fig9} we compare fRG results for $G(V_g)$
to NRG data. Two different parameter sets with purely local interaction
$U$, $\Gamma_1^L=\Gamma_1^R=\Gamma/2$, and 
$\varepsilon_{1,\sigma}= \varepsilon_{2,\sigma} =V_g$ are used. 
In Fig.~\ref{fig8} the parameters are in the spin-singlet 
regime: $U/\Gamma=8$, $t_{1,2}/\Gamma=4$.\cite{Cornaglia} 
The fRG and NRG data agree well. From the gate voltage dependence 
of the many-body eigenvalues  of the double dot system without 
the leads\cite{Zitko} it becomes clear that each of the plateau-like 
resonances centered around $\pm V_{g}^{p}$ 
can be understood as the resonance in the single dot 
case, with the difference that the approximate width  of the two peaks is 
$\Delta=U/2+2t_{1,2} \left[ 1- \sqrt{1+(U/4t_{1,2})^2}\right]$
instead of $U$. Around $V_g=0$ both dots are half-filled and a 
local spin-singlet is formed leading to a suppression of the 
conductance. The parameters 
$U/\Gamma=2$ and $t_{1,2}/\Gamma=0.2$ of Fig.~\ref{fig9} are taken 
from the two-stage Kondo regime.\cite{Cornaglia}  As $U/\Gamma$ is moderate
the first stage of the Kondo effect is not well pronounced which explains 
that the conductance away from half-filling (that is for $\left| V_g
\right| \geq U/2$) vanishes only slowly with increasing 
$\left| V_g \right|$. For these gate voltages the fRG and
NRG data are almost indistinguishable. The second stage of the Kondo
effect generates a dip at $\omega=0$ in the spectral function of 
dot $j=1$ at half filling,\cite{Cornaglia} that leads to the 
suppression of $G$ near $V_g=0$. For small $t_{1,2}$ the 
scale $U/t^2_{1,2}$ relevant in
the second stage of the Kondo effect becomes very large and, not 
surprisingly, our approximation deviates quantitatively (not
qualitatively) from the NRG data
for $\left| V_g \right|\leq U/2$. For these gate
voltages some of the flowing interactions become large and our
truncation scheme is questionable. Within our approximation 
the parameter regime of very large $U/t^2_{1,2}$, in which in addition 
to the two-stage Kondo effect a  Fano-like feature in $G(V_g)$ was 
observed,\cite{Zitko} is out of reach. 

In Figs.~\ref{fig8} and \ref{fig9}, in addition to $G$ we show the 
dot occupancies $\left<n_j\right>$, with $n_j = n_{j,\uparrow} + 
n_{j,\downarrow}$, computed by fRG. Because of the large inter-dot
hopping in Fig.~\ref{fig8} the occupancies behave similarly for 
all $V_g$. 
For small $t_{1,2}$ (Fig.~\ref{fig9}) the side-coupled 
dot $j=2$ is coupled very weakly to the leads and the particle number 
of this dot changes very abruptly across the transmission resonances.  
Close to $V_g=0$ the occupancies $\left<n_j\right>$ depend 
non-monotonically on $V_g$. This follows from an interaction-induced 
non-monotonicity of the effective level positions. 
Similar behavior was observed in a spin-polarized parallel double-dot 
model.\cite{Sindel,Koenig,cir} In contrast the total 
dot charge $\left< n \right>=\left< n_1 \right> + \left< n_2 \right>$ 
is monotonic. Similar to the single-dot case it is directly related to the
conductance\cite{Cornaglia} 
\begin{eqnarray}
\label{Gn}
G/(e^2/h)=2 \sin^2{(\left< n \right> \pi
  /2)} \, .
\end{eqnarray}
The (spin-independent) transmission phase $\alpha_\sigma$ 
for the parameter sets of Figs.~\ref{fig8} and \ref{fig9} is shown in
Fig.~\ref{fig10}. Experimentally this phase is accessible if the
side-coupled double dot is placed in one arm of a ``two-path'' 
Aharonov-Bohm interferometer and the current oscillations as a function of a
magnetic flux enclosed by the two arms are measured. Such measurements
were already performed for 
single multi-level dots.\cite{HeiblumExps} Across a transmission
resonance the phase continuously changes by $\pi$. Associated to
$G(V_g=0)=0$ is a phase jump of $\pi$.  
From $\alpha_\sigma$ the conductance can be computed as  
$G/(e^2/h)=2 \sin^2{\alpha_\sigma}$. Note that
this relation together with Eq.~(\ref{Gn}) only implies 
$\alpha_\sigma = \left( \left< n_\sigma \right> +m \right) \pi$, with $m \in
{\mathbb Z}$ and the observed phase jumps by $\pi$ are not excluded. 
The weak shoulder at $\alpha_\sigma \approx \pi /2$ for the 
parameters of Fig.~\ref{fig8} is related to the Kondo effect and 
develops into a plateau if $U/\Gamma$ is increased. 

\begin{figure}[tb]
\begin{center}
\includegraphics[width=0.45\textwidth,clip]{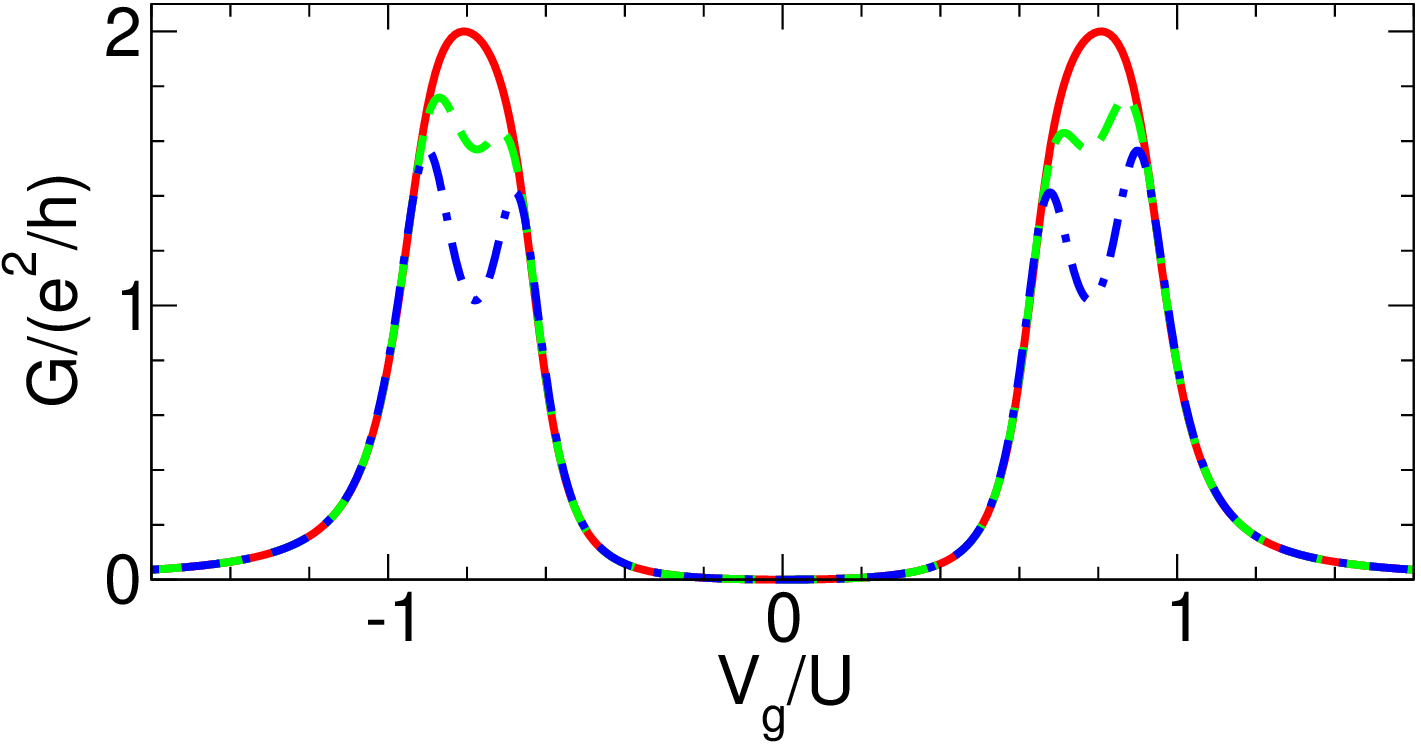}
\end{center}
\vspace{-0.6cm}
\caption[]{(Color online) 
Gate Voltage dependence of the total conductance $G$ for the
parameters of Fig.~\ref{fig8} and different magnetic fields 
${\mathcal H}=0$ (solid line), ${\mathcal H}=0.065\Gamma=0.5 T_K$ 
(dashed line), 
${\mathcal H}=0.13\Gamma=T_K$ (dashed-dotted line). 
\label{fig11}}
\begin{center}
\includegraphics[width=0.45\textwidth,clip]{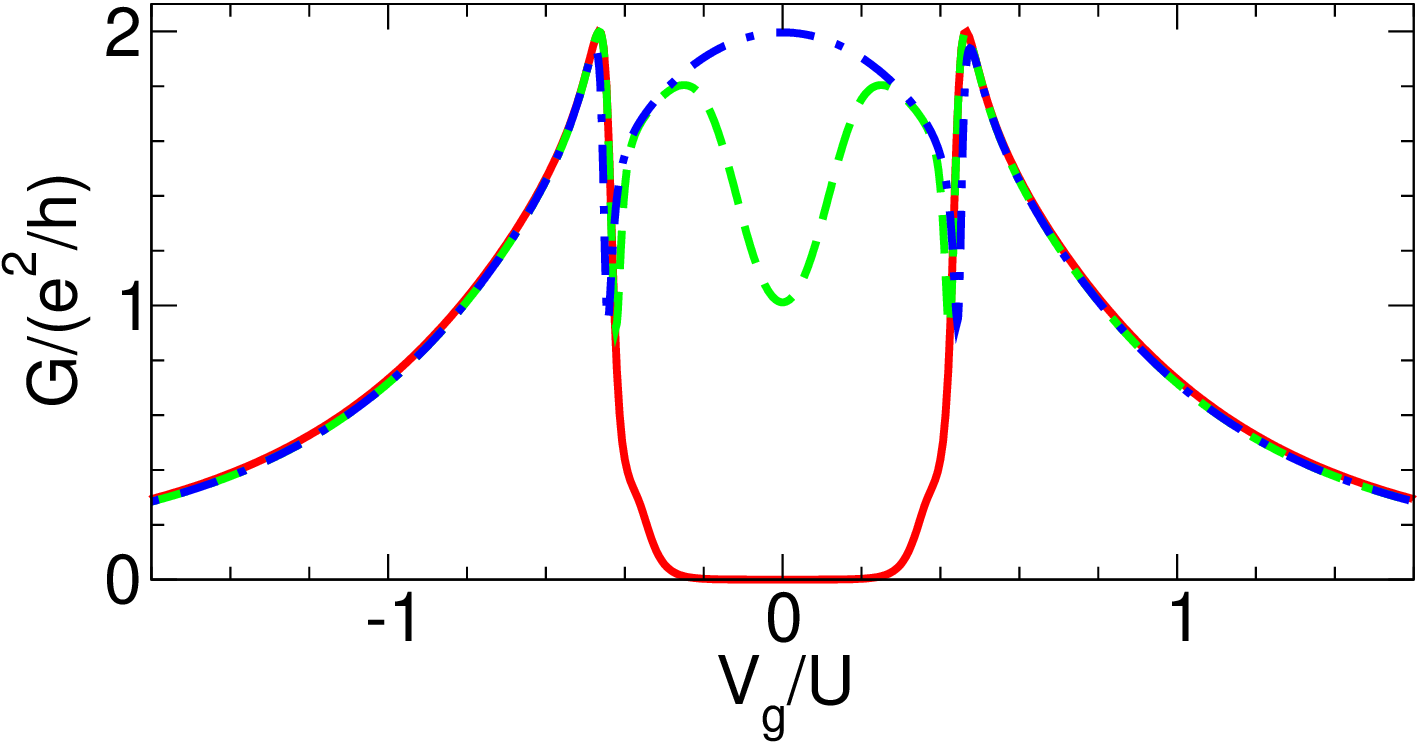}
\end{center}
\vspace{-0.6cm}
\caption[]{(Color online) 
Gate Voltage dependence of the total conductance $G$ for the
parameters of Fig.~\ref{fig9} and different magnetic fields 
${\mathcal H}=0$ (solid line), ${\mathcal H}=0.003 \Gamma=T^0=
0.0026 T_K^0$ 
(dashed line), ${\mathcal H}=0.03\Gamma= 10 T^0 = 0.026 T_K^0$ 
(dashed-dotted line).   
\label{fig12}}
\end{figure}

From the $\mathcal H=0$ conductance curves the two regimes (local
spin-singlet and two-stage Kondo) cannot be distinguished
unambiguously. In particular, in both cases $G$ is suppressed very
strongly close to half-filling. The $G(V_g)$ curves in the two
regimes change quite differently in the presence of a magnetic field
that lifts the spin-degeneracy of both levels: 
$\varepsilon_{1,\uparrow}=\varepsilon_{2,\uparrow}=V_g-{\mathcal
  H}/2$,  $\varepsilon_{1,\downarrow}=\varepsilon_{2,\downarrow}
=V_g+{\mathcal H}/2$. 
As for the single dot case studying the magnetic field dependence 
also enables us to define Kondo temperatures. 
For the two parameter sets of Figs.~\ref{fig8}
and \ref{fig9} $G(V_g)$ for different ${\mathcal H}$ is shown 
in Figs.~\ref{fig11} and \ref{fig12}. 
In the two-stage Kondo regime we define a scale $T_K^0$ 
for the first stage of the Kondo effect considering the single dot 
case, i.e.~setting 
$t_{1,2}=0$. For the parameters of Fig.~\ref{fig12} we obtain
$T_K^0/\Gamma=1.16$ in good agreement with the NRG estimate
$T_K^0/\Gamma=1$.\cite{Cornaglia} 
Because of its very small energy scale $T_0 \ll T_K^0$ the second 
stage of the Kondo effect gets destroyed already 
by a tiny magnetic field ${\mathcal H} \ll T_K^0$ and the local 
minimum of $G$ at half-filling evolves into a maximum as shown 
in Fig.~\ref{fig12}. We define $T_0$ by the magnetic field required 
to obtain $G(U,V_g=0,{\mathcal  H}=T_0)=e^2/h$. For the parameters 
of Fig.~\ref{fig12} this leads to $T_0/\Gamma=0.003$. 
For larger $t_{1,2}$ (spin-singlet regime), 
as shown in Fig.~\ref{fig11}, the spin-singlet is
the ground state up to moderate $\mathcal H$ and $V_g=0$ remains a 
minimum. The magnetic field dependence of the two resonances is
similar to the single dot case.
In this parameter regime it is meaningful to define a Kondo
temperature $T_K$ by $G(U,V_g=\pm V_g^p,{\mathcal H}=T_K)=e^2/h$. 
For the parameters of Fig.~\ref{fig11} we obtain $T_K/\Gamma=0.13$,
which is close to the NRG result $T_K/\Gamma=0.156$.\cite{Cornaglia}
In both parameter regimes the evolution of $G(V_g)$ with $\mathcal H$ 
is similar to the one with $T$ discussed in Ref.~\onlinecite{Cornaglia}. 

For the present setup we also studied the role of
(a) asymmetric dot-lead couplings $\Gamma_1^L  \neq  \Gamma_1^R$, (b) 
asymmetric local interactions $U_{1,1}^{\uparrow,\downarrow} 
\neq U_{2,2}^{\uparrow,\downarrow}$, and (c) a nearest-neighbor 
interaction $U_{j,\bar j}^{\sigma,\sigma'}$, with $\bar j$ denoting
the complement of $j$. In case
(a) the overall height of $G$ decreases with increasing asymmetry while
the line shape remains invariant similar to the case of a single
dot.\cite{Meir} Asymmetric local interactions and 
nearest-neighbor interactions only lead to quantitative changes but no new
physics.   

\section{Parallel double dots}
\label{paralleldoubledot}

As our most complex example we finally study the parallel double dot
of Fig.~\ref{fig1} E. The large number of parameters allows
for a variety of different regimes and we here refrain from giving a
complete account of the physics of this setup. 
Instead, we focus on a few cases to exemplify that our approximation 
scheme can be used to uncover interesting effects. A more detailed 
discussion  will be given in an upcoming publication.
We here only consider the case without an inter-dot
hopping. If present initially such a term can always be tuned away
by a basis transformation on bonding and anti-bonding dot states. 
Note that in the noninteracting case the double dot can equally be 
viewed as a single two-level dot. Including the interaction both cases
might be distinguished by the choice of non-vanishing 
matrix elements (see below).
Correlated multi-level dots have recently attracted much interested
in connection with the puzzling observation of ``universal'' phase
lapses in a series of measurements of the transmission phase through 
quantum dots.\cite{HeiblumExps} 
The conductance for the present setup
with full spin polarization and small level detuning was earlier
studied using the simplest fRG-based approximation scheme (considering
only the flow of the self-energy) and NRG. In the absence of spin-Kondo
physics the inter-dot (inter-level) interaction leads to novel 
correlation-induced resonances located exponentially (in $U$) close to
$V_g=0$.\cite{cir}  

The noninteracting projected ``effective Hamiltonian'' of the present
setup is given by   
\begin{eqnarray}
\label{effh0}
h_\sigma(i\omega) = \left( \begin{array}{cc} 
\varepsilon_{1,\sigma} - i \Gamma_1  \sgn(\omega)& 
- i \gamma \sgn(\omega)\\
- i \gamma^\ast \sgn(\omega) &
\varepsilon_{2,\sigma}  - i \Gamma_2 \sgn(\omega) 
  \end{array} \right) 
\end{eqnarray}
with 
\begin{eqnarray}
\label{defdef}
\Gamma_j = \sum_l \Gamma_j^l \; , \;\;\; \;\;\; \gamma =  
\sqrt{\Gamma_1^L \Gamma_2^L} + e^{i \phi} \sqrt{\Gamma_1^R
  \Gamma_2^R} \; . 
\end{eqnarray}
We furthermore define
\begin{eqnarray}
\label{defdefdef}
\Gamma = \sum_j \Gamma_j = \sum_j \sum_l \Gamma_j^l \; .
\end{eqnarray}
Here $\phi$ denotes a magnetic flux that goes through
the ring geometry. If the system is viewed as a single two-level dot,
$\phi$ is restricted to $0$ and $\pi$, on which we will focus in the
following. We denote the relative sign of the hopping matrix elements
$t_j^l$ by $s= e^{i \phi} = \pm 1$. 
The noninteracting propagator $ {\mathcal G}_0(i\omega)$ follows by 
the matrix inversion Eq.~(\ref{matrixinversion}). For generic
parameters during the fRG flow
Eqs.~(\ref{finalflowsigma}) and (\ref{finalflowgamma}) the level
positions are renormalized and a (real) direct inter-dot hopping is
generated. Applying the projection technique Eq.~(\ref{projection})
the conductance can be computed from the matrix elements of the 
projected Green function at
$\Lambda=0$ 
\begin{widetext}
\begin{eqnarray}
\label{parallelG}
G_\sigma(V_g) = 4 \, \frac{e^2}{h} \, \left|  \sqrt{\Gamma_1^L \Gamma_1^R}
   {\mathcal G}^{\Lambda=0}_{\sigma;1,1}(0) + \sqrt{\Gamma_2^L \Gamma_1^R} 
   {\mathcal G}^{\Lambda=0}_{\sigma;1,2}(0) +  s \sqrt{\Gamma_1^L \Gamma_2^R}
   {\mathcal G}^{\Lambda=0}_{\sigma;2,1}(0) + s \sqrt{\Gamma_2^L \Gamma_2^R}
   {\mathcal G}^{\Lambda=0}_{\sigma;2,2}(0) \right|^2 \; .
\end{eqnarray}   
The large number of parameters make it  essential to analyze 
the conductance for the noninteracting case before considering 
the effect of two-particle interactions. 
%For the  other setups 
%studied (Figs.~1 A-D) the noninteracting physics is less rich 
%and we did not discuss it in detail. 
%For the present double dot a
A 
closed expression for $G_\sigma(V_g)$ at $U_{j,j'}^{\sigma,\sigma'}=0$ 
can be obtained from Eq.~(\ref{parallelG}) by replacing 
$ {\mathcal G}^{\Lambda=0}(0)$ by $ {\mathcal G}_0(0)$ 
\begin{eqnarray}
\label{U0conduct}
G_\sigma(V_g) = \frac{e^2}{h} \; \frac{4 \left[\Gamma_1^L \Gamma_1^R
  \varepsilon^2_{2,\sigma}+ \Gamma_2^L \Gamma_2^R
  \varepsilon^2_{1,\sigma} + 2 s \sqrt{\Gamma_1^L \Gamma_1^R 
\Gamma_2^L \Gamma_2^R}  \varepsilon_{1,\sigma}
\varepsilon_{2,\sigma}\right]}{\left[\Gamma_1^L \Gamma_2^R + \Gamma_2^L
\Gamma_1^R  - 2 s \sqrt{\Gamma_1^L \Gamma_1^R 
\Gamma_2^L \Gamma_2^R}- \varepsilon_{1,\sigma}
\varepsilon_{2,\sigma}\right]^2 +\left[  \varepsilon_{1,\sigma} \Gamma_2 +
\varepsilon_{2,\sigma} \Gamma_1 \right]^2 } \; .
\end{eqnarray} 
\end{widetext}
We here focus on the spin-degenerate case with ${\mathcal H}=0$. For
generic level-lead couplings\cite{cir} 
$\Gamma_j^l$ and $\delta=  \varepsilon_{2,\sigma} -
\varepsilon_{1,\sigma} \geq 0$ the gate voltage dependence 
($ \varepsilon_{1/2,\sigma} = \mp \delta/2 +V_g$) of
Eq.~(\ref{U0conduct}) is characterized by two peaks (of height 
$\leq e^2/h$) and a conductance zero. The latter follows from 
perfect destructive interference at a particular $V_g$. Associated to 
the zero is a jump of the transmission phase by $\pi$. Further 
details of $G_\sigma(V_g)$ depend on $s$. For $s=+1$ the conductance 
zero is located between the two conductance peaks for all $\delta$. 
For $\delta \to 0$ the peak positions depend on the asymmetry of 
the $\Gamma_j^l$ and the separation of the peaks is small if the 
$\Gamma_j^l$ are almost equal. For equal $\Gamma_j^l$   
and $\delta=0$ the zero disappears. For $s=-1$ and fixed $\Gamma_j^l$
the position of the conductance zero with respect to the peaks 
depends on $\delta$. For $\delta \to 0$ it is located between 
the two conductance peaks, while it lies
outside for large $\delta$. In the crossover regime
between these limiting cases one of the peaks vanishes, while the
other becomes broader and splits up into two resonances separated by a
minimum with non-vanishing conductance. For $\Gamma_j^l$ with 1-2
symmetry and $\delta \to 0$ the conductance vanishes for all $V_g$ 
as the model can be mapped onto a model with two levels, each one only 
coupled to one of the leads. 
In this (and only this) case a pseudo-spin variable 
is conserved leading to an orbital Kondo effect if a nearest-neighbor 
interaction is 
added.\cite{Boese1,Boese2} For fixed, asymmetric $\Gamma_j^l$ and
$\delta \to 0$ the separation of the two conductance peaks for $s=-1$
is much larger than for $s=+1$. For certain classes of parameters,
some of them we mentioned already (e.g., equal $\Gamma_j^l$ and
$\delta=0$; see Ref.~\onlinecite{cir} for details), $G(V_g)$ does not follow
the behavior described above. Here, we do not investigate these
special cases but focus on the generic behavior.  

\begin{figure}[tb]
\begin{center}
\includegraphics[width=0.45\textwidth,clip]{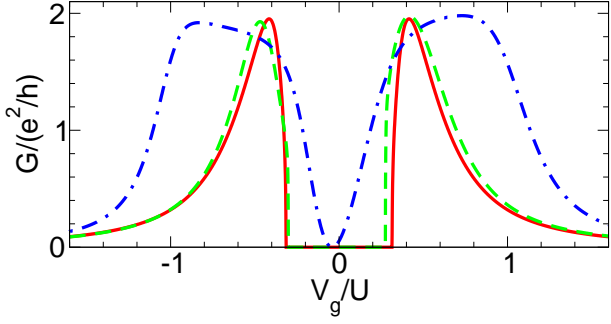}
\end{center}
\vspace{-0.6cm}
\caption[]{(Color online) 
Gate Voltage dependence of the total conductance $G$ of a parallel 
double dot (see Fig.~\ref{fig1} E) with generic $\Gamma_j^l$, spin-degenerate
levels, purely local interactions, and different level detunings $\delta$. 
The parameters are $U/\Gamma=4$,
$\Gamma_1^L/\Gamma=0.27$, $\Gamma_1^R/\Gamma=0.33$,  
$\Gamma_2^L/\Gamma=0.16$, $\Gamma_2^R/\Gamma=0.24$, $s=1$,
$\delta/\Gamma=0$ (solid line), $\delta/\Gamma=1$ (dashed line), and 
$\delta/\Gamma=5$ (dashed-dotted line). \label{fig13}} 
\end{figure}

In Fig.~\ref{fig13} we show $G(V_g)$ for a generic set of $\Gamma_j^l$,
$s=+1$, and different $\delta$. We assume a purely local interaction 
$U$, which might experimentally be realized if the setup is implemented
as two single-level dots of sufficient spatial separation.
Close to half-filling the conductance is strongly suppressed. In
particular this holds for small $\delta$. We note that around $V_g=0$
the flowing two-particle vertex becomes large which indicates that our
approximation scheme becomes less accurate. Further down we return to
this issue.  For $\delta/\Gamma \leq 1$ the curves of 
Fig.~\ref{fig13} are similar to the one of Fig.~\ref{fig9} obtained 
for the  side-coupled double dot with small inter-dot hopping 
$t_{1,2}$. We argue that 
this similarity follows from a relation between the two double-dot 
geometries which we investigate next. 

Taking the even and odd linear combination of dot states the
noninteracting side-coupled dot (index $s$) with 
$\left( t_1^L \right)_s$, $\left( t_1^R\right)_s$, 
$\left( t_{1,2} \right)_s$, and $\left( \varepsilon_{j,\sigma}
\right)_s 
= V_g$ 
can be mapped on the parallel dot with parameters 
\begin{eqnarray}
\label{map}
t_1^L & = & \left( t_1^L \right)_s/\sqrt{2} \; , \; \; \; 
t_1^R=\left( t_2^R \right)_s/\sqrt{2} \; , \nonumber \\
t_2^L & = & \left( t_1^L \right)_s/\sqrt{2} \; , \; \; \;  
t_2^R=\left( t_2^R \right)_s/\sqrt{2}  \; , \nonumber \\ 
\varepsilon_{1,\sigma} & = & V_g -  \left( t_{1,2}  \right)_s \; , \; \; \;
\varepsilon_{2,\sigma}  =  V_g  +  \left( t_{1,2}  \right)_s \; , 
\end{eqnarray}
that is, $\delta=2 \left( t_{1,2}  \right)_s$. Note that small
$\left(t_{1,2}\right)_s$ correspond to small $\delta$.
This subset of the large parameter space of the parallel double dot is
characterized by a 1-2 symmetry of the hoppings and $s= 1$. A local 
interaction of the side-coupled dot 
\begin{eqnarray}
\label{sdot}
U_s \sum_{j} n^s_{j,\uparrow}  n^s_{j,\downarrow} \; .   
\end{eqnarray}
maps onto
\begin{eqnarray}
\label{trafdot}
&& \frac{U_s}{2} \left( \sum_j n_{j, \uparrow}
  n_{j,\downarrow} +  n_{1, \uparrow}   n_{2, \downarrow} 
  + n_{1, \downarrow} n_{2, \uparrow}     \right) \nonumber \\
&& +  \frac{U_s}{2} \left(d_{2, \uparrow}^\dag d_{1, \uparrow} d_{2,
    \downarrow}^\dag  d_{1, \downarrow}  + d_{2, \uparrow}^\dag 
     d_{1, \uparrow} d_{1, \downarrow}^\dag  d_{2, \downarrow}
   \right. \nonumber \\
&& \left. + d_{1, \uparrow}^\dag d_{2, \uparrow} d_{2, \downarrow}^\dag d_{1,
  \downarrow} +  d_{1, \uparrow}^\dag d_{2, \uparrow} d_{1,
  \downarrow}^\dag d_{2, \downarrow}
\right) 
\end{eqnarray}
for the parallel double dot.  Besides the usual density interactions
also correlated hoppings are generated by this 
transformation.\cite{footnote} 

These considerations show that the cases studied in 
Figs.~\ref{fig9} and \ref{fig13} (for $\delta/\Gamma \leq 1$) 
are related but not completely equivalent as in Fig.~\ref{fig13} we only
considered a local interaction and no 1-2 symmetry of the
$\Gamma_j^l$.  Slowly increasing the amplitudes of the additional 
interactions Eq.~(\ref{trafdot}) and tuning the $\Gamma_j^l$ towards
1-2 symmetry we convinced ourselves that both the symmetry and the
additional interactions do not play an essential role and the results
for the parallel double dot with $\delta/\Gamma \leq 1$ 
of Fig.~\ref{fig13} can be
interpreted as for the side-coupled double dot in the two-stage Kondo
regime.  The strong
suppression of the conductance of the parallel double dot 
around half-filling for small $\delta$ 
then follows from the second stage of the Kondo effect. From the 
side-coupled dot we already know that our approximation scheme 
overestimates this suppression and we believe that the same holds for 
the present geometry. Note that for $\delta=0$, $G(V_g)$ is 
symmetric around $V_g=0$ as the Hamiltonian is symmetric under a 
particle-hole transformation if in addition $V_g \to -V_g$. 
Despite the asymmetry of the $\Gamma_j^l$ the two resonances have
almost unitary height. For small $\delta$ the occupancies of the two 
levels are similar, sharply change across the two resonances, and
display a plateau at 1 close to $V_g=0$. The transmission phase 
increases by $\pi$ across the resonances and jumps by $\pi$ in 
between the peaks.  

For large $\delta$, $G(V_g)$ in Fig.~\ref{fig13} can be understood 
from the eigenvalues of the many-body dot Hamiltonian without the 
leads, in analogy to the limit of large $\left(t_{1,2}\right)_s$ 
for the side-coupled dot. Each of the resonances then 
corresponds to a single-level 
resonance, as discussed in Sec.~\ref{singledot}. Because of the 
asymmetry of the $\Gamma_j^l$, the conductance does not reach the 
unitary limit and the two resonances have different line shape.  
Associated with the resonances is a change of one of the 
$\left< n_{j} \right>$ from 2 to 0 (change of the phase from 0 
to $\pi$) and a plateau at 1 (at $\pi/2$ for the phase) within 
the resonance. For the $\Gamma_j^l$ of Fig.~\ref{fig13}, 
$\left< n_2 \right>$ changes across the peak at $V_g<0$ and 
$\left< n_1 \right>$ across the one at $V_g>0$. The plateaus in 
$\left< n_j \right>$ and $\alpha_\sigma$ are due to the Kondo effect 
that is active at each of the resonances. At the conductance zero 
between the conductance peaks the transmission phase jumps by $\pi$. 

\begin{figure}[tb]
\begin{center}
\includegraphics[width=0.45\textwidth,clip]{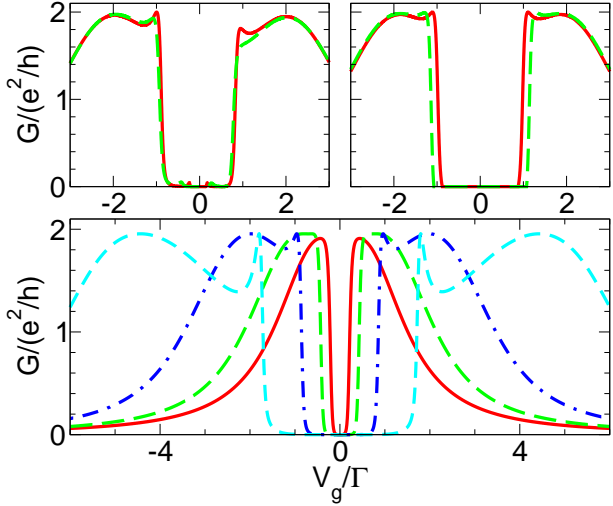}
\end{center}
\vspace{-0.6cm}
\caption[]{(Color online) 
Gate Voltage dependence of the total conductance $G$ of a parallel 
double dot with generic $\Gamma_j^l$, spin-degenerate
levels, and (almost) equal local and nearest-neighbor density
interactions. 
The parameters are $\Gamma_1^L/\Gamma=0.5$, $\Gamma_1^R/\Gamma=0.25$,  
$\Gamma_2^L/\Gamma=0.07$, $\Gamma_2^R/\Gamma=0.18$, and $s=1$.
{\it Lower panel:} Equal level positions on both dots 
$\delta=0$ and equal local and nearest-neighbor interactions 
$U$ with $U/\Gamma=0.5$, (solid line) $1$
(long dashed line), $2$ (dashed-dotted line), and $4$ (short dashed
line).  {\it Upper left panel:} 
Stability of the novel resonances for non-vanishing 
level detuning $\delta$ at $U/\Gamma=2$. $\delta/\Gamma=0.03$ (solid line) and 
$\delta/\Gamma=0.07$ (dashed line).  {\it Upper right panel:}  
Stability of the novel 
resonances for asymmetric interactions at $\delta=0$. 
$U_{1,1}^{\sigma,\bar \sigma}/\Gamma = 1.9$, 
$U_{2,2}^{\sigma,\bar \sigma}/\Gamma = 2.1$,
$U_{j,\bar j}^{\sigma,\sigma}/\Gamma = 1.8$, $U_{j,\bar j}^{\sigma,\bar
  \sigma}/\Gamma =  2$ (solid line) and 
$U_{1,1}^{\sigma,\bar \sigma}/\Gamma = 1.8$, 
$U_{2,2}^{\sigma,\bar \sigma}/\Gamma = 2.2$,
$U_{j,\bar j}^{\sigma,\sigma}/\Gamma = 1.7$, $U_{j,\bar j}^{\sigma,\bar
  \sigma}/\Gamma =  2.2$ (dashed line).  Note that here $V_g$ is
shown in units of $\Gamma$. 
 \label{fig14}} 
\end{figure}

Scanning the parameter space with fixed $s=+1$ we encountered an
interesting correlation effect for a model with  
almost equal local and nearest-neighbor density interactions, 
small $\delta$, and generic $\Gamma_j^l$. It is exemplified in
Fig.~\ref{fig14}. In the lower panel of Fig.~\ref{fig14} we show
$G(V_g)$ for $\delta=0$ and $U_{j,j'}^{\sigma,\sigma'} = U
(1-\delta_{j,j'} \delta_{\sigma,\sigma'})$. Increasing $U$, the peak
conductance increases, the two maxima become flat, and for $U$ larger
than a $\Gamma_j^l$ dependent critical $U_c(\{\Gamma_j^l\})$ both
split up into two peaks. For $U>U_c$ the height of all four resonances
is equal and $U$ independent. The two outermost peaks are located at
$V_g \approx \pm U$ and the inner two at $V_g \approx \pm U/2$. This
scenario is quite similar to the appearance of additional 
correlation induced resonances in a spin-polarized model of a 
parallel double dot with nearest-neighbor interaction as discussed 
in Ref.~\onlinecite{cir}. The novel resonances follow from the
combined presence of correlations and quantum interference. In the 
upper panels of Fig.~\ref{fig14} we show that the effect is robust 
against a small level detuning $\delta$ and a small asymmetry of the
$U_{j,j'}^{\sigma,\sigma'}$. It will be further investigated in an
upcoming publication. One can expect that in experimentally realized 
single two-level dots the inter- and intra-dot density interactions
are indeed (almost) equal and the additional resonances should thus be
observable in such systems. 

We next study the case $s=-1$. As for $s=+1$ above we first
investigate systems with purely local interactions 
(two well-separated single-level 
dots) and further down a situation with (almost) equal local and
nearest-neighbor interactions (a single two-level dot). 

\begin{figure}[tb]
\begin{center}
\includegraphics[width=0.45\textwidth,clip]{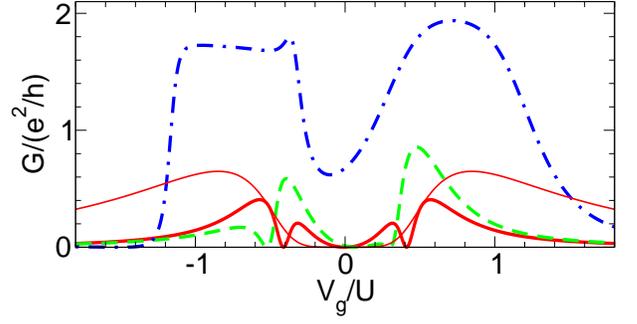}
\end{center}
\vspace{-0.6cm}
\caption[]{(Color online) 
Gate Voltage dependence of the total conductance $G$ of a parallel 
double dot with generic $\Gamma_j^l$, spin-degenerate
levels, purely local interactions, and different level detunings $\delta$. 
The parameters are $U/\Gamma=4$ (thick lines), $U/\Gamma=1$ (thin line)
$\Gamma_1^L/\Gamma=0.5$, $\Gamma_1^R/\Gamma=0.35$,  
$\Gamma_2^L/\Gamma=0.05$, $\Gamma_2^R/\Gamma=0.1$, $s=-1$,
$\delta/\Gamma=0$ (solid line), $\delta/\Gamma=0.6$ (dashed line), and 
$\delta/\Gamma=6$ (dashed-dotted line). \label{fig15}} 
\end{figure}

In Fig.~\ref{fig15} we show $G(V_g)$ for $s=-1$, generic $\Gamma_j^l$, a
purely local interaction, and different $\delta$. Without loss of
generality we consider $\Gamma_j^l$ with a fairly large 1-2 
asymmetry. Otherwise the conductance at small $\delta$ is very 
small. This follows from the  fact that for 1-2 symmetric $\Gamma_j^l$ 
and $\delta=0$, $G(V_g) =0$ for all $V_g$, as discussed above. The
thin solid curve shows the case with $\delta=0$ and $U/\Gamma=1$, that is
qualitatively similar to the $U=0$ case. Increasing $U$ (at $\delta=0$) 
each of the two peaks splits up into two with a conductance zero in 
between. In Fig.~\ref{fig15} this is shown as the thick solid line 
for $U/\Gamma=4$. The dashed and dashed-dotted lines show how the
conductance evolves with increasing level detuning $\delta$ at fixed 
$U/\Gamma=4$. The conductance zeros either vanish or are pushed
towards larger $|V_g|$. The overall conductance increases. For large
$\delta$ the two remaining resonances can be understood as in the single-dot, 
single-level case. Across the resonance centered around $-\delta/2$
($+\delta/2$) level 2 (1) gets filled. As the $\Gamma_2^l$ are small 
the effective interaction $U/(\Gamma_2^L+ \Gamma_2^R)$ relevant for 
the Kondo effect is large and the resonance at $V_g<0$ has a well-developed 
plateau-like line shape. These results can most easily be 
understood by making contact with the case of left-right symmetric 
$\Gamma_j^l$. In this case the off-diagonal element $\gamma$ 
[see Eq.~(\ref{defdef})] of the non-interacting ``effective Hamiltonian'' 
(\ref{effh0}) vanishes. As the interaction is purely local no such 
term (that is a hopping between $j=1$ and $j=2$) is generated during 
the fRG flow and the double-dot problem separates into two single-dot 
problems. In particular, ${\mathcal  G}_{\sigma;1,2}^{\Lambda=0}$ 
vanishes and the conductance Eq.~(\ref{parallelG}) is given by the sum 
of two single-level Green functions. Breaking of the left-right
symmetry leads to a small initial off-diagonal element, that stays
small during the fRG flow. For that reason also the results of
Fig.~\ref{fig15} can be understood from the single-level case. 
E.g., this explains the position $V_g \approx \pm \delta/2$ of the
resonances at large $\delta$. A more detailed discussion will be given
elsewhere.   
    
\begin{figure}[tb]
\begin{center}
\includegraphics[width=0.45\textwidth,clip]{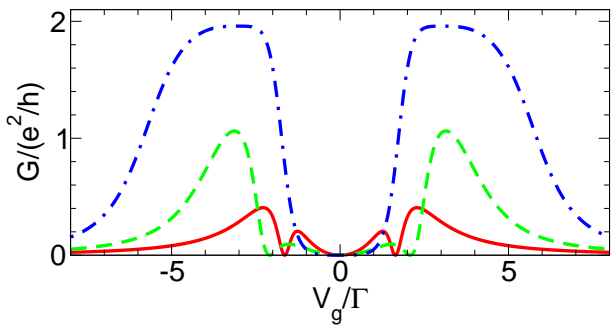}
\end{center}
\vspace{-0.6cm}
\caption[]{(Color online) 
Gate Voltage dependence of the total conductance $G$ of a parallel 
double dot with generic $\Gamma_j^l$, $s=-1$, spin-degenerate
levels, $\delta=0$, local interaction $U/\Gamma=4$, and different
nearest-neighbor interactions $U'=0$ (solid line), $U'/\Gamma=2$ 
(dashed line), and $U'/\Gamma=4$ (dashed-dotted line). 
The $\Gamma_j^l$ are as in Fig.~\ref{fig15}. Note that here $V_g$ is
shown in units of $\Gamma$. 
\label{fig16}}
\end{figure}

In Fig.~\ref{fig16} we show how the four-peak structure obtained for a 
sufficiently large local interaction $U$ at
$\delta=0$ and $s=-1$ disappears in the presence of a nearest-neighbor 
interaction $U'$ (single two-level dot). Starting out with four peaks 
of almost equal height at 
$U'=0$, for increasing $U'$ the two outermost peaks gain weight, while 
the inner two peaks lose weight. At $U'=U$ only two resonances
remain.

\section{Summary and outlook}
\label{summary}

In the present paper we have introduced a novel approximate method to
study electronic transport through systems of quantum dots with local
Coulomb correlations. Our fRG-based approximation scheme results in a 
set of coupled differential flow equations for the interaction-dependent 
effective level positions and 
inter-level hoppings as well as the interaction matrix elements. 
We have solved the flow equations for a variety of dot systems and
computed the linear conductance, the transmission phase, and level
occupancies. 

For several setups and parameter sets we compared our
results to existing exact Bethe ansatz and high precision NRG data. 
For a single dot the agreement is excellent up to the largest Coulomb
interaction for which Bethe ansatz or NRG data are available. In
particular, we showed analytically that our approach covers the
signatures of the Kondo effect. The method performs very well also in
the presence of a magnetic field lifting the spin degeneracy and in
case of an additional direct transmission channel between the leads
(Fano effect). The field dependence of $G(V_g)$ allows us to extract 
the Kondo temperature that compares well to the exact expression. 
For linear chains of dots (short Hubbard chains) we
found quantitative agreement with NRG data up to fairly large
interactions. We studied the effect of a left-right asymmetry of the
level-lead couplings that leads to the surprising vanishing of
resonances if sufficiently large. For a side-coupled double-dot system
we demonstrated that our method can be used to describe both the 
``molecular'' spin-singlet regime at large inter-dot hoppings
$t_{1,2}$ and the two-stage Kondo regime at small $t_{1,2}$. For 
not too small $t_{1,2}$ and too large interactions the approximate
results agree well with NRG data. Our
approximation scheme becomes questionable if $U/t_{1,2}^2$ becomes too
large. We showed that the occupancies in the two-stage Kondo regime
depend non-monotonically of the gate voltage. At $T=0$ the different
physics of the two regimes can be probed by a magnetic field lifting
the spin degeneracy. Finally we demonstrated that the physics of a
parallel single-level double dot (single two-level dot) is very
rich. For a certain parameter regime we showed that $G(V_g)$ can be
understood in analogy to the side-coupled double dot. For other
classes of parameters we surprisingly found additional resonance
peaks that should be observable experimentally in existing double-dot 
setups (two-level dots). A more 
complete analysis of the transport properties of the
parallel double dot (single two-level dot)
will be given in an upcoming publication. 

The present paper shows that our approximation scheme is a 
simple-to-implement and reliable method to study transport through 
dot systems with a significant number of interacting
degrees of freedom. It can be used to study setups that due to their
complexity cannot be tackled by NRG. Even in problems that can
be investigated by NRG our method is far superior in terms of the numerical 
effort required, e.g.\ enabling efficient analysis of parameter
dependencies. 

In summary, we have shown that the fRG based approximation scheme is a
very promising tool to investigate the interesting physics of quantum 
dot systems resulting from Coulomb correlations (Kondo effect)
and quantum interference. 

\section*{Acknowledgments}
We are grateful to P.~Cornaglia, T.~Costi, J.~von Delft, W.~Hofstetter, and
A.~Oguri for providing their NRG and Bethe ansatz data and thank 
S.~Andergassen, J.~von Delft, R.~Hedden, 
W.~Hofstetter, W.~Metzner, Th.~Pruschke, M.~Reginatto,  
H.~Schoeller, K.~Sch\"onhammer, and M.~Weyrauch for valuable discussions.  
T.E.~is grateful for support 
by a Feoder Lynen fellowship of the Alexander von Humboldt foundation 
and the Istituto Nazionale di Fisica della Materia--SMC--CNR.  
V.M.~is grateful to the Deutsche Forschungsgemeinschaft (SFB 602) for support.


\begin{thebibliography}{*}
\bibitem{Sohn} \textit{Mesoscopic Electron Transport}, edited by
  L.L.~Sohn, L.P.~Kouwenhoven, and G.~Sch\"on (Kluwer, Dordrecht,
  1997).
\bibitem{Loss} D.~Loss and D.P.~DiVincenzo, Phys.\ Rev.\ A {\bf 57},
  120 (1998); D.~Loss and E.V.~Sukhorukov, Phys.\ Rev.\ Lett.\ {\bf
    84}, 1035  (2000).
\bibitem{Florian} F.~Marquardt and C.~Bruder, Phys.\ Rev.\ B {\bf 68}, 
195305 (2003).
\bibitem{Holleitner1} A.W.~Holleitner, C.R.~Decker, H.~Qin, K.~Eberl,
  and R.H.~Blick,  Phys.\ Rev.\ Lett.\ {\bf 87}, 256802 (2001);
A.~W.~Holleitner, R.H.~Blick, A.K.~H\"uttel, K.~Eberl, and
J.P.~Kotthaus, Science {\bf 297}, 70 (2002).
\bibitem{Chen} J.C.~Chen, A.M.~Chang, and M.R.~Melloch, Phys.\ Rev.\ Lett.\ 
  {\bf 92}, 176801 (2004).
\bibitem{Sigrist} M.~Sigrist, A.~Fuhrer, T.~Ihn, K.~Ensslin,
  S.E.~Ulloa, W.~Wegschneider, and M.~Bichler, 
Phys.\ Rev.\ Lett.\ {\bf  93}, 066802 (2004).
\bibitem{Holleitner2} A.~W.~Holleitner, A.~Chudnovskiy, D.~Pfannkuche,
  K.~Eberl, and R.H.~Blick,  Phys.\ Rev.\ B {\bf 70}, 075204 (2004). 
\bibitem{Petta1} J.R.~Petta, A.C.~Johnson, C.M.~Marcus, M.P.~Hanson,
  and A.C.~Gossard, Phys.\ Rev.\ Lett.\ {\bf 93}, 
186802 (2004).
\bibitem{Craig} N.J.~Craig, J.M.~Taylor, E.A.~Lester, C.M.~Marcus,
  M.P.~Hanson, and A.C.~Gossard, Science {\bf 304}, 565 (2004). 
\bibitem{Koppens} F.H.L.~Koppens, J.A.~Folk, J.M.~Elzerman, R.~Hanson,
  L.H.~Willems van Beveren, I.T.~Vink, H.P.~Tranitz, W.~Wegscheider,
  L.P.~Kouwenhoven, and L.M.K.~Vandersypen, Science {\bf 309}, 
1346 (2005).
\bibitem{Petta2} J.R.~Petta, A.C.~Johnson, J.M.~Taylor, E.A.~Laird,
  A.~Yacoby, M.D.~Lukin, C.M.~Marcus, M.P.~Hanson, and A.C.~Gossard, 
  Science {\bf 309}, 2180 (2005).
\bibitem{Johnson} A.C.~Johnson, J.R.~Petta, J.M.~Taylor, A.~Yacoby,
  M.D.~Lukin, C.M.~Marcus, M.P.~Hanson, and A.C.~Gossard, Nature {\bf 435}, 925
  (2005). 
\bibitem{Hewson} A.C.~Hewson, \textit{The Kondo Problem to Heavy
    Fermions}, (Cambridge University Press, Cambridge, UK, 1993). 
\bibitem{Glazman} L.~Glazman and M.~Raikh, JETP Lett.\ {\bf 47}, 452 (1988).
\bibitem{Ng} T.~Ng and P.~Lee, Phys.\ Rev.\ Lett.\ {\bf 61}, 1768 (1988).
\bibitem{Goldhaber} D.~Goldhaber-Gordon, H.~Shtrikman, D.~Mahalu,
  D.~Abusch-Magder, U.~Meirav, and M.A.~Kastner, Nature {\bf
    391}, 156 (1998).
\bibitem{Wiel} W.~van der Wiel, S.~De Franceschi, T.~ Fujisawa,
  J.M.~Elzerman, S.~Tarucha, and L.P.~Kouwenhoven, Science {\bf
    289}, 2105 (2000).
\bibitem{Glazmanrev} L.P.~Kouwenhoven and L.~Glazman, Physics World, 33
  (2001).   
\bibitem{Tsvelik} A.M.~Tsvelik and P.B.~Wiegmann, Adv.\ Phys.\ {\bf
    32}, 453 (1983).
\bibitem{Theo1} T.A.~Costi, A.C.~Hewson, and V.~Zlatic, J.~Phys.:
  Condens.\ Matter {\bf 6}, 2519 (1994).
\bibitem{Gerland} U.~Gerland, J.~von Delft, T.A.~Costi, and Y.~Oreg,
 Phys.\ Rev.\ Lett.~{\bf  84}, 3710 (2000). 
\bibitem{Oguri2} A.~Oguri and A.C.~Hewson, J.~Phys.\ Soc.\ Jpn.\ {\bf
    74}, 988 (2005); A.~Oguri, Y.~Nisikawa, and A.C.~Hewson, 
    {\it ibid} {\bf 74}, 2554 (2005).
\bibitem{Oguri3} Y.~Nisikawa and A.~Oguri, Phys.~Rev.~B {\bf 73}, 
125108 (2006). 
\bibitem{Zitko2} R.~\v{Z}itko, J.~Bon\v{c}a, A.~Ram\v{s}ak, and 
T.~Rejec, Phys.~Rev.~B {\bf 73}, 153307 (2006). 
\bibitem{Fano} U.~Fano, Phys.\ Rev.\ {\bf 124}, 1866 (1961). 
\bibitem{Hofstetterkondofano} W.~Hofstetter, J.~K\"onig, and
  H.~Schoeller, Phys.\ Rev.\ Lett.\ {\bf 87}, 156803 (2001). 
\bibitem{Kim1} T.-S.~Kim and S.~Hershfield, Phys.\ Rev.\ B {\bf 63},
  245326 (2001). 
\bibitem{Cornaglia} P.S.~Cornaglia and D.R.~Grempel, Phys.\ Rev.\ B
  {\bf 71}, 075305 (2005).
\bibitem{Zitko} R.~\v{Z}itko and J.~Bon\v{c}a, Phys.~Rev.~B {\bf 73}, 035332
  (2006). 
\bibitem{Izumida} W.~Izumida, O.~Sakai, and Y.~Shimizu,
  J.~Phys.\ Soc.\ Jpn.\ {\bf 66}, 717 (1997). 
\bibitem{Boese1} D.~Boese, W.~Hofstetter, and H.~Schoeller, 
Phys.\ Rev.\ B {\bf 64}, 125309 (2001). 
\bibitem{Boese2} D.~Boese, W.~Hofstetter, and H.~Schoeller, 
Phys.\ Rev.\ B {\bf 66}, 125315 (2002). 
\bibitem{cir} V.~Meden and F.~Marquardt, Phys.~Rev.~Lett.~{\bf 96}, 
146801 (2006). 
\bibitem{HeiblumExps} A.~Yacoby, M.~Heiblum, D.~Mahalu, and
  H.~Shtrikman, Phys.\ Rev.\ Lett.\ {\bf 74}, 4047 (1995); 
R. Schuster, E.~ Buks, M.~Heiblum, D.~Mahalu, V.~Umansky,
and H.~Shtrikman, Nature {\bf 385}, 417 (1997);
M. Avinun-Kalish,  M.~Heiblum, O.~Zarchin, D.~Mahalu, and 
V.~Umansky, Nature {\bf 436}, 529 (2005).
\bibitem{Wilson} K.G.~Wilson, Rev.~Mod.~Phys.\ {\bf 47}, 773 (1975).
\bibitem{Krishnamurthy} H.B.~Krishnamurthy, J.W.~Wilkins, and
  K.G.~Wilson, Phys.\ Rev.\ B {\bf 21}, 1044 (1980). 
\bibitem{Kashcheyevs} V.~Kashcheyevs, A.~Aharony, and
  O.~Entin-Wohlman, Phys.~Rev.~B {\bf 73}, 125338 (2006). 
\bibitem{Oreg} Y.~Oreg and Y.~Gefen, Phys.\ Rev.\ B {\bf 55}, 13726
  (1997). 
\bibitem{Herbert} H.~Schoeller in \textit{Low-Dimensional Systems}, 
  ed.\ T.~Brandes (Springer, 1999), p.~137. 
\bibitem{Buesser} See C.A.~B\"usser, G.B.~Martin, K.A.~Al-Hassanieh,
  A.~Moreo, and E.~Dagotto, Phys.\ Rev.\ B {\bf 70}, 245303 (2004) and 
references therein.
\bibitem{Ogurikritik} A.~Oguri, cond-mat/0310139.
\bibitem{Torio} M.E.~Torio, K.~Hallberg, A.H.~Ceccatto, and
  C.R.~Proetto, Phys.\ Rev.\ B {\bf 65}, 085302 (2002). 
\bibitem{Kotliar} G.~Kotliar and A.E.~Ruckenstein,
  Phys.\ Rev.\ Lett.\ {\bf 57}, 1362 (1986).
\bibitem{Dong1} B.~Dong and X.L.~Lei, J.~Phys.: Condens.\ Matter
  {\bf 13}, 9245 (2001).
\bibitem{Takahashi} J.~Takahashi and S.~Tasaki, cond-mat/0603337. 
\bibitem{Dong2} B.~Dong and X.L.~Lei, Phys.\ Rev.\ B {\bf 63},
  235306 (2001). 
 \bibitem{Tanaka} See Y.~Tanaka and N.~Kawakami, Phys.\ Rev.\ B {\bf 72},
  085304 (2005) and references therein.
\bibitem{GS} O.~Gunnarsson and
  K.~Sch\"onhammer, Phys.\ Rev.\ B {\bf 31}, 4815 (1985). 
\bibitem{Rejec} T.~Rejec and A.~Ram\v{s}ak, Phys.\ Rev.\ B {\bf 68},
  035342 (2003).
\bibitem{Manfred} M.~Salmhofer, \textit{Renormalization}, (Springer, Berlin, 1998).
\bibitem{2dsystems} 
 D.~Zanchi and H.J.~Schulz, Phys.\ Rev.\ B {\bf 61}, 13609 (2000);
 C.J.~Halboth and W.~Metzner, {\it ibid} {\bf 61}, 7364 (2000);
 C.~Honerkamp, M.~Salmhofer, N.~Furukawa, and T.M.~Rice, 
 {\it ibid} {\bf 63}, 035109 (2001). 
\bibitem{Ralf} R.~Hedden, V.~Meden, Th.~Pruschke, and
  K.~Sch\"on\-hammer, J.~Phys.: Condens.\ Matter {\bf 16}, 5279 (2004).
\bibitem{Hon} C.~Honerkamp, D.~Rohe, S.~Andergassen, and T.~Enss, 
  Phys.\ Rev.\ B {\bf 70}, 235115 (2004).
\bibitem{Wetterich} C.~Wetterich, Phys.\ Lett.\ B {\bf 301}, 90 (1993). 
\bibitem{Morris} T.R.~Morris, Int.\ J. Mod.\ Phys.\ A {\bf 9}, 2411 (1994).
\bibitem{SalmhoferHonerkamp} M.~Salmhofer and C.~Honerkamp, 
  Prog.\ Theor.\ Phys.\ {\bf 105}, 1 (2001). 
\bibitem{STV} S.~Andergassen, T.~Enss, and V.~Meden, 
Phys.\ Rev.\ B {\bf 73} 153308 (2006).
\bibitem{footnoteminus} Note that in contrast to Eq.~(\ref{leadham}),
  Eq.~(\ref{directham}) does not contain a minus sign. We here follow the
  convention of Ref.~\onlinecite{Hofstetterkondofano}.  
\bibitem{Datta} For a review on the Landauer-B\"uttiker approach to
  transport see: S.~Datta, \textit{Electronic Transport in Mesoscopic
    Systems}, (Cambridge University Press, Cambridge, 1995). 
\bibitem{Taylor} J.R.~Taylor, \textit{Scattering Theory}, (John Wiley and
  Sons, New York, 1972).
\bibitem{Oguri1} A.~Oguri, J.~Phys.\ Soc.\ Jpn.\ {\bf 70}, 2666 (2001).
\bibitem{NegeleOrland} J.W.~Negele and H.~Orland, 
  \textit{Quantum Many-Particle Physics}, (Addison-Wesley, Reading, 1988).
\bibitem{lecturenotes} V.~Meden, lecture notes on the ``Functional
  renormalization group'',
  http://www.theorie.physik.uni-goettingen.de/$\sim$meden/funRG/
\bibitem{Meir} Y.~Meir and N.S.~Wingreen, Phys.\ Rev.\ Lett.\ {\bf 68}, 
2512 (1992). 
\bibitem{EnssThesis} T.~Enss,
  Ph.D. thesis, Universit\"at Stuttgart 2005,
%  urn:nbn:de:bsz:93-opus-22587,
  URL: http://elib.uni-stuttgart.de/opus/volltexte/2005/2258, cond-mat/0504703.
\bibitem{Theo2} T.A.~Costi, Phys.\ Rev.\ B {\bf 64}, 241310(R) (2001).
\bibitem{footnotekatanin} Note that in Ref.~\onlinecite{Ralf} in
  addition to the scheme involving the single-scale propagator
  $S^\Lambda$ a modified approximation is used. In the latter, $S^\Lambda$ in
  the flow equation for $\Gamma^\Lambda$ is replaced by
  $-\frac{\partial}{\partial \Lambda} {\mathcal G}^\Lambda$. We
  verified that for the truncation  scheme used in the present
  publication this modification does not lead to 
  any significant improvement of the results. This has to be
  contrasted to the truncation scheme of Ref.~\onlinecite{Ralf}
  involving frequency dependence in which the above replacement 
  leads to much better agreement with NRG results.  
\bibitem{footnotephases} Note that the transmission phase
  $\alpha_\sigma$ is given by the sum
  of the phases $\delta_{\text even}$ and $\delta_{\text odd}$ defined in
  Ref.~\onlinecite{Oguri3}.
\bibitem{Sindel} M.~Sindel, A.~Silva, Y.~Oreg, and J.~von Delft,
  Phys.\ Rev.\ B {\bf 72}, 125316
(2005).
\bibitem{Koenig} J.~K\"onig and Y.~Gefen, Phys.\ Rev.\ B {\bf 71},
  201308(R) (2005).
\bibitem{footnote} Note that Eq.~(\ref{trafdot}) does not 
contain terms proportional to $n_{1,\sigma} n_{2,\sigma}$. 
If a nearest-neighbor interaction of strength $U_s$ is added 
the interacting part of the side-coupled double-dot 
Hamiltonian reads $\frac{U_s}{2} \sum_{(j,\sigma) \neq (j',\sigma')} 
n^s_{j,\sigma}  n^s_{j',\sigma'}$. This interaction is then 
invariant under the basis transformation onto the parallel 
double dot.\cite{Boese2}  

\end{thebibliography}
\end{document}